\documentclass[twocolumn,table,xcdraw]{bytedance_seed}

\usepackage[toc,page,header]{appendix}

\usepackage{minitoc}
\usepackage{tikz}
\usepackage{amsmath}

\usepackage{filecontents}
\usepackage{hyperref}
\usepackage{multirow}
\usepackage{amsthm}
\usepackage{graphicx}
\usepackage{amsfonts}
\usepackage{xcolor}

\usepackage[linesnumbered,noend]{algorithm2e}
\SetAlgoHangIndent{0pt}
 \usepackage{url}
\usepackage{tikz}
\newcommand*\circled[1]{\tikz[baseline=(char.base)]{
            \node[shape=circle,draw,inner sep=0.2pt] (char) {#1};}}

\newcommand{\paper}{SwiftSpec}

\title{ \paper: Ultra-Low Latency LLM Decoding by Scaling Asynchronous Speculative Decoding  }

\author[1,2]{Ziyi Zhang}
\author[1\dagger]{Ziheng Jiang}
\author[1]{Chengquan Jiang}
\author[1]{Menghan Yu}
\author[1]{Size Zheng}
\author[1]{Haibin Lin}
\author[2]{Henry Hoffmann}
\author[1, \dagger]{Xin Liu}

\affiliation[1]{ByteDance Seed}
\affiliation[2]{University of Chicago}

\contribution[\dagger]{Corresponding authors}

\abstract{

Low-latency decoding for large language models (LLMs) is crucial for applications like chatbots and code assistants, yet generating long outputs remains slow in single-query settings. Prior work on speculative decoding (which combines a small \emph{draft} model with a larger \emph{target} model) and tensor parallelism has each accelerated decoding.  However, conventional approaches fail to apply both simultaneously due to imbalanced compute requirements (between draft and target models), KV-cache inconsistencies, and communication overheads under small-batch tensor-parallelism.

This paper introduces \paper{}, a system that targets ultra-low latency for LLM decoding. \paper{} redesigns the speculative decoding pipeline in an asynchronous and disaggregated manner, so that each component can be scaled flexibly and remove draft overhead from the critical path. To realize this design, \paper{} proposes parallel tree generation, tree-aware KV cache management, and fused, latency-optimized kernels to overcome the challenges listed above. Across 5 model families and 6 datasets, \paper{} achieves an average of $1.75\times$ speedup over state-of-the-art speculative decoding systems and, as a highlight, serves Llama3-70B at 348 tokens/s on 8 Nvidia Hopper GPUs, making it the fastest known system for low-latency LLM serving at this scale.
}

\date{\today}
\correspondence{Ziheng Jiang at \email{ziheng.jiang@bytedance.com}, Xin Liu at \email{liuxin.ai@bytedance.com}}

\begin{document}
\maketitle

%不需要目录就注释掉 注意目录不要和第一页放在一块 要有\newpage
%\newpage
%\tableofcontents
%\newpage

% \input{sections/introduction}
% \input{sections/relatedwork}
% \input{sections/approach}
% \input{sections/experiments}
\section{Introduction}

\begin{figure*}[t]
    \centering
    \includegraphics[width=\linewidth]{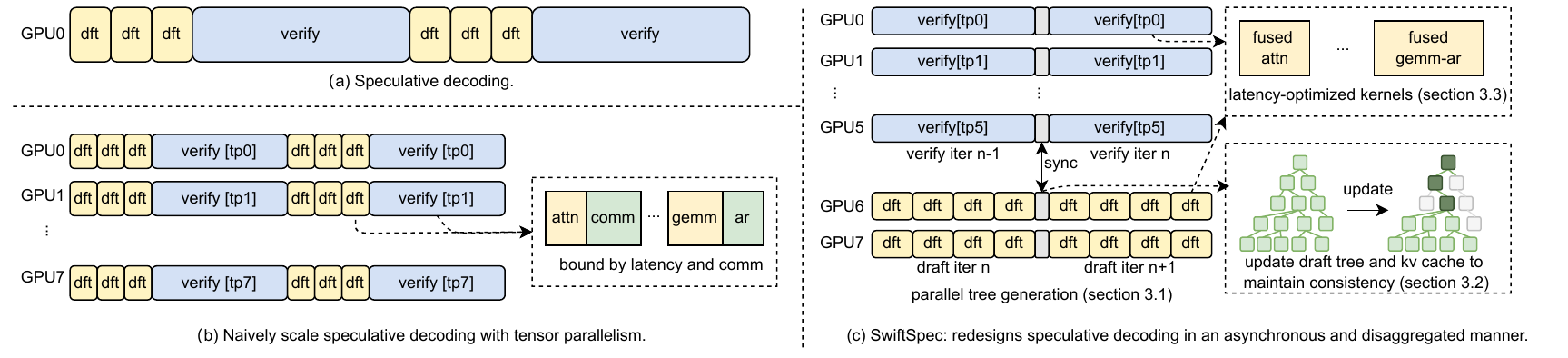}
    % \vspace{-0.2in}
    \caption{Overview of speculative decoding variants. (a) Conventional speculative decoding with sequential draft/verify steps. (b) Tensor-parallel decoding with reduced latency but communication overhead and GPU under utilization. (c) \paper{}: Our approach, combining parallel tree generation (\S\ref{sec:parellel_tree_generation}), KV cache consistency (\S\ref{sec:kv_cache_management}), and latency-optimized kernels (\S\ref{sec:low_latency_kernels}). 
    }
    \label{fig:sys_arch}
    % \vspace{-0.2in}
    % \vspace{-20pt}
\end{figure*}

The remarkable capacity of large language models (LLMs) to learn from vast datasets has been instrumental in enabling the rapid proliferation of emerging applications across diverse domains, including chatbots \cite{hariri2025unlockingpotentialchatgptcomprehensive}, search \cite{search1_xiong2024searchengineservicesmeet, search2_caramancion2024largelanguagemodelsvs}, and personalized recommendation systems \cite{hariri2025unlockingpotentialchatgptcomprehensive, yang2023palrpersonalizationawarellms}. 
Real-time applications, such as interactive code assistants \cite{izadi2024languagemodelscodecompletion, coder1guo2023longcoderlongrangepretrainedlanguage, coder2chen2021evaluatinglargelanguagemodels} and robotics \cite{robot1hou2025codeasmonitorconstraintawarevisualprogramming, robot2}, impose stringent limits on models' decoding latency. Recently, Chain-of-thought (CoT)  \cite{wei2023chainofthoughtpromptingelicitsreasoning, deepseekai2025deepseekr1incentivizingreasoningcapability} is also increasingly adopted to improve reasoning quality. A single CoT inference can decode tens of thousands of tokens and take more than ten minutes to complete. Reducing decoding latency is, therefore, critical.

In LLM-serving systems, there is an inherent trade-off between throughput and latency under the same compute resources. This work investigates how to achieve ultra-low latency decoding in single-request scenarios, where existing serving frameworks---designed to maximize throughput under service-level-objective (SLO) constraints---often fall short. For instance, a 4-bit quantized LLama-70B model running on 8 NVIDIA H800 GPUs can take approximately 30s to generate a response of about 3000 tokens when deployed with the popular LLM serving frameworks.

\begin{table}[t]
\resizebox{\columnwidth}{!}{%
\begin{tabular}{c|cccc}
           & \# gpus=1 & \# gpus=2 & \# gpus=4 & \# gpus=8 \\ \hline
Llama3-1B  & 1.58ms  & 1.58ms  & 1.73ms  & 1.72ms  \\
Llama3-3B  & 2.77ms  & 2.61ms  & 2.80ms  & 3.27ms  \\
Llama3-8B  & 4.30ms   & 3.46ms  & 3.50ms   & 3.78ms  \\
Llama3-70B & 24.78ms & 15.90ms & 11.86ms & 11.22ms
\end{tabular}%
}
\caption{Time per inference for 4 different int4 AWQ models under Llama3 family under different degrees (ngpus) of tensor parallelism under a batch size of 8  }
% \vspace{-0.4in}
\label{tab:infer_time_intro}
\end{table}

\textit{Speculative decoding} \cite{cai2024medusasimplellminference, Miao_2024, chen2023acceleratinglargelanguagemodel, leviathan2023fastinferencetransformersspeculative} accelerates LLM inference in single-request scenarios. 
Speculative decoding consists of two distinct phases: a draft phase followed by a verification phase. During the draft phase, a relatively small draft model rapidly generates a sequence of candidate tokens (and, in some variants, a tree-structured set of candidates). During the subsequent verification phase, a significantly larger target model validates all candidates by performing a batch inference, thereby emitting multiple tokens at once and reducing decoding latency. Prior work typically treats the draft and verification phases as strictly sequential operations because of their data dependencies \cite{Miao_2024, cai2024medusasimplellminference, li2024eagle2fasterinferencelanguage}, as shown in Figure \ref{fig:sys_arch}(a). This design places the draft phase on the critical path as an additional overhead, preventing speculative decoding from fully realizing its latency-reduction potential.

\textit{Tensor parallelism}~\cite{shoeybi2020megatronlmtrainingmultibillionparameter} is another technique to reduce the decoding latency by scaling the computation resources. Tensor parallelism partitions the model weights across multiple GPUs and then performs all-reduce operations to aggregate the partial results. However, a straightforward combination of tensor parallelism with speculative decoding, as shown in Figure \ref{fig:sys_arch}(b), is ineffective. In speculative decoding, the draft and target models are co-located on the same devices. Because the two models differ greatly in size, applying the same degree of tensor-parallelism to both cannot yield optimal system latency. The smaller draft model reaches the point of diminishing returns sooner: once its weights are already finely sharded, further increasing the tensor-parallelism no longer reduces latency, because other overheads---most notably inter-GPU communication---dominate. As Table \ref{tab:infer_time_intro} shows, under a batch size of 8. extending from 2 GPUs to 4 GPUs, a lightweight Llama 3B draft model has even longer inference time (increasing from 2.61ms to 2.80ms per inference), while the Llama 70B target model still has good improvements (decreasing from 15.90ms to 11.86ms).

To effectively combine speculative decoding with tensor parallelism and achieve ultra-low decoding latency, we redesign the speculative decoding process in an asynchronous, disaggregated manner (Figure \ref{fig:sys_arch}(c)). We partition GPUs into two groups: verification and draft. Rather than co-located on the same hardware, the target model runs on the verification group, and the draft model runs on the draft group. The verification and draft phases proceed in parallel: while the verification group verifies iteration $n-1$, the draft group concurrently produces candidates for iteration $n$. When a verification iteration is complete, the verification group synchronizes the validated tokens with the draft group and obtains the next set of candidate tokens to be verified. Under this design, (1) the draft and target models can be flexibly scaled to different degrees of parallelism, and (2) the dependencies between the two phases are decoupled, removing the draft phase from the critical path.

Realizing this design poses three system-level challenges. First, while the verification group is still performing parallel validation and has not yet obtained a definitive answer for the current iteration, the draft group must still generate the candidate set for the next iteration. Second, maintaining key-value cache consistency between complex drafting models (e.g., tree-structured draft models) and the target model is non-trivial. When tree-based draft generation runs in parallel, newly accepted tokens may force the draft model to discard invalid branches. It is important (yet challenging) to keep a consistent view of the KV cache of accepted tokens and the draft tokens that might be useful in the future. Third, hiding communication latency during decoding is challenging. For example, when draft and target models are under tensor parallelism, it is hard to overlap the all-reduce operation with other operations since they usually remain on the critical path. Furthermore, the GPU kernels, usually optimized for higher throughput, have suboptimal performance under low batch sizes, spending most of the time on the latency of data movement and kernel launch.

We present~\textit{\paper{}}, a novel system that achieves ultra-low decoding latency for LLMs (Figure \ref{fig:sys_arch}(c)), significantly reducing the decoding latency in single-request scenarios. To address the above challenges, \paper{} introduces: (a) \textbf{Parallel tree generation}. We allocate the draft and target models onto different sets of GPUs, eliminating inter-dependencies and allowing each model to generate tokens or verify them independently. While the target model verifies one batch, the draft model simultaneously produces future candidate tokens, ensuring high GPU utilization. This allows us to scale each model according to its own compute requirements. (b) \textbf{Consistent KV cache management}. After each verification step, we carefully reorganize the KV Cache of both the draft and target models to maintain consistency. For the draft model, we develop a scheme to keep the accepted and future tokens consistent with the draft tree, even when some guesses are incorrect and some part of the draft tree is invalidated. This approach also maximizes the reuse of the previously computed KV cache values. (c) \textbf{Latency-optimized kernels}. We develop latency-optimized kernels that minimize synchronization barriers and unnecessary data transfers, accelerating inference in low-batch scenarios. Using the Nvidia Collective Communication Library's Low Latency (NCCL LL) protocol, we develop a \emph{fused} GEMM with all-reduce and an attention operator without any explicit synchronization barriers. Furthermore, we fuse the multiple operations in the Switched Gated Linear Unit (SwiGLU) operator to decrease latency.

We evaluate \paper{} using five different model families and six different datasets. \paper{} consistently outperforms the baselines, achieving an average of $1.75\times$ decoding speed using 8 GPUs over the best baseline. As a highlight, \paper{} serves Llama3-70B with an average decoding speed of 348 output tokens per second, currently the fastest solution we are aware of using Nvidia Hopper GPUs. 
% \pagebreak

Our contributions are:
\begin{itemize}
\item We identify the scalability challenges of speculative decoding under tensor parallelism in existing LLM serving systems.
\item We present \paper{}, which integrates techniques including parallel tree generation, consistent KV cache management, and latency-optimized kernels to redesign speculative decoding in an asynchronous, disaggregated manner.
\item We conduct a comprehensive evaluation of \paper{} across five model families and six benchmark datasets, in which \paper{} consistently outperforms the baselines.

\end{itemize}

\section{Background and System-level Challenges}

\begin{table*}[]
\resizebox{\linewidth}{!}{%
\begin{tabular}{cccccc}
\hline

\multicolumn{1}{c}{\textbf{System}} &
  \multicolumn{1}{c}{\textbf{Spec Type}} &
  \multicolumn{1}{c}{\textbf{Spec Parallel}} &
  \multicolumn{1}{c}{\textbf{Consistent KV-Cache Management}} &
  \multicolumn{1}{c}{\textbf{Latency-Optimized Kernels}} \\ \hline

\textbf{SpecInfer, EAGLE \cite{Miao_2024, li2024eagle2fasterinferencelanguage}} &
  \cellcolor[HTML]{FFFFFF}{\color[HTML]{009901} Tree} &
    \cellcolor[HTML]{FFFFFF}{\color[HTML]{FE0000} No} &
  \cellcolor[HTML]{FFFFFF}{\color[HTML]{009901} easy to manage} &
  \cellcolor[HTML]{FFFFFF}{\color[HTML]{FE0000} standard kernels} \\
\textbf{PEARL \cite{liu2024parallelspeculativedecodingadaptive}} &
  \cellcolor[HTML]{FFFFFF}{\color[HTML]{FE0000} Sequence} &
    \cellcolor[HTML]{FFFFFF}{\color[HTML]{009901} Yes} &
  \cellcolor[HTML]{FFFFFF}{\color[HTML]{009901} easy to manage} &
  \cellcolor[HTML]{FFFFFF}{\color[HTML]{FE0000} standard kernels} \\
\textbf{PipeInfer \cite{butler2024pipeinferacceleratingllminference}} &
  \cellcolor[HTML]{FFFFFF}{\color[HTML]{009901} Tree} &
    \cellcolor[HTML]{FFFFFF}{\color[HTML]{009901} Yes} &
    \cellcolor[HTML]{FFFFFF}{\color[HTML]{FE0000} no fine-grained re-use of draft cache} &
  \cellcolor[HTML]{FFFFFF}{\color[HTML]{FE0000} standard kernels/CPU operations} \\
\textbf{AMUSD \cite{mcdanel2024amusdasynchronousmultidevicespeculative}} &
  \cellcolor[HTML]{FFFFFF}{\color[HTML]{FE0000} Sequence} &
    \cellcolor[HTML]{FFFFFF}{\color[HTML]{009901} Yes} &
  \cellcolor[HTML]{FFFFFF}{\color[HTML]{009901} easy to manage} &
  \cellcolor[HTML]{FFFFFF}{\color[HTML]{FE0000} standard kernels} \\
\textbf{SGLang, vLLM \cite{zheng2024sglangefficientexecutionstructured, vllmkwon2023efficientmemorymanagementlarge}} &
  \cellcolor[HTML]{FFFFFF}{\color[HTML]{009901} Tree} &
    \cellcolor[HTML]{FFFFFF}{\color[HTML]{FE0000} No} &
  \cellcolor[HTML]{FFFFFF}{\color[HTML]{009901} easy to manage} &
  \cellcolor[HTML]{FFFFFF}{\color[HTML]{FE0000} limited optimization} \\
\textbf{\paper{} (Ours)} &
  \cellcolor[HTML]{FFFFFF}{\color[HTML]{009901} \textbf{Tree}} &
    \cellcolor[HTML]{FFFFFF}{\color[HTML]{009901} \textbf{Yes}} &
  \cellcolor[HTML]{FFFFFF}{\color[HTML]{009901} \textbf{fined-grained reorganization, zero waste}} &
  \multicolumn{1}{c}{\cellcolor[HTML]{FFFFFF}{\color[HTML]{009901} \textbf{fused latency-optimized ops}}} \\ \hline
\end{tabular}%
}
\caption{The comparison between \paper{} and other existing baselines}
\label{tab:motivation_baseline_comparison}
\vspace{-0.15in}
\end{table*}

\subsection{Parallelism for LLM Decoding}
LLM inference (e.g., \cite{vaswani2023attentionneedtransformer, touvron2023llamaopenefficientfoundation, openai2024gpt4technicalreport}) is typically accelerated through model parallelism, which distributes computation across multiple GPUs. The two most common strategies are intra-operator parallelism (such as tensor parallelism, which splits matrix multiplications across GPUs) \cite{shoeybi2020megatronlmtrainingmultibillionparameter} and inter-operator parallelism (such as pipeline parallelism, which distributes layers across devices) \cite{huang2019gpipeefficienttraininggiant, fan2020dapplepipelineddataparallel}.

In single-request serving, tensor parallelism (TP) is generally preferred. By splitting large matrix operations across GPUs and aggregating results via all-reduce, TP reduces compute latency. Pipeline parallelism, while useful for high-throughput scenarios, often adds per-token latency due to sequential layer execution.

However, TP alone is insufficient to achieve low latency in speculative decoding. Draft models are typically small and gain relatively little from TP, while large target models benefit more from increased TP degrees. These asymmetric scaling needs pose challenges in resource allocation and utilization. \paper{} directly  addresses these issues through parallel tree generation (\S\ref{sec:parellel_tree_generation}) and latency-optimized kernels (\S\ref{sec:low_latency_kernels}).

\subsection{GPU Constraints for Low Batch Size}
Modern GPUs feature high-bandwidth memory (HBM) and fast interconnects (e.g., NVLink), and they are typically optimized for large-batch throughput workloads. In small-batch inference, common in interactive LLM serving, GPU performance suffers due to underutilized kernels and communication overhead.

In transformer layers executed with TP, three operators dominate runtime: GEMM, attention, and all-reduce. Under small batch sizes ($\leq 8$), these operators exhibit poor bandwidth and compute utilization due to short execution and frequent synchronization. Some targeted optimizations, such as low-bit quantization \cite{lin2024awqactivationawareweightquantization, xiao2024smoothquantaccurateefficientposttraining}, reduce communication costs but fall short of a holistic solution.

Similarly, state-of-the-art LLM serving frameworks have reduced communication costs for small sizes by using the Nvidia collective communication library all-reduce operator \cite{zheng2024sglangefficientexecutionstructured, vllmkwon2023efficientmemorymanagementlarge}.  Again, this falls short of a holistic solution.  \paper{} instead introduces latency-optimized kernels (\S\ref{sec:low_latency_kernels}) that fuse GEMM and all-reduce operators to greatly reduce overhead and enable fine-grained communication.

\subsection{Tree-based Speculative Decoding}
\label{sec:motivation_spec}

%The LLM inference process typically consists of two phases: prefill (processing the initial prompt) and decoding (iteratively predicting each subsequent token). Since decoding usually consumes most of the generation time, our work focuses on accelerating this phase of LLM serving.

\subsubsection{Overview of Tree-based Approaches} Speculative decoding methods leverage a smaller \emph{draft model} to accelerate inference from a larger \emph{target model} \cite{cai2024medusasimplellminference, Miao_2024, chen2023acceleratinglargelanguagemodel, leviathan2023fastinferencetransformersspeculative}. These techniques typically proceed in two stages: a drafting phase and a verification phase. In the drafting phase, the draft model iteratively proposes a short sequence or a partial tree \cite{Miao_2024} of potential next tokens. In the subsequent verification phase, the target model verifies these proposed tokens by performing a batched forward pass over the drafted tokens and then sampling from the results. The draft model then proceeds to predict the next set of tokens, and this process is repeated until the target model concludes generation.

Conventional speculative decoding approaches impose dependencies between the target model and the draft model. Parallel sequence-based speculative decoding partially addresses this challenge by concurrently running the draft model to generate the next guesses (under the assumption that previous guesses are correct) while the target model verifies those preceding guesses \cite{liu2024parallelspeculativedecodingadaptive, mcdanel2024amusdasynchronousmultidevicespeculative}. 

However, sequence-based methods typically exhibit lower-quality guesses compared to tree-based approaches. 
In tree-based speculative decoding, the draft model proposes a tree of candidate token paths instead of a linear sequence. The target model verifies a path through this tree. This approach tends to achieve higher \emph{compression ratios}, i.e., tokens verified per target inference. Compression ratio directly impacts the end-to-end latency, as higher ratios reduce the number of target model invocations.

So, while tree-based methods can improve compression ratios, they also impose more complexity on the overall decoding system. First, while sequential approaches often generate the single most probable next token, it remains unclear how to optimally construct tree nodes in parallel to maximize the compression ratio. Second, managing the KV cache to maintain a consistent view of accepted tokens and potentially useful draft tree nodes introduces further complexity. \paper{} supports tree-based parallel speculative decoding by addressing these two limitations through two key innovations: parallel tree generation and consistency management in asynchronous decoding (\S\ref{sec:parellel_tree_generation}--\ref{sec:kv_cache_management}).

\label{sec:motivation_motivation}
\begin{figure}
    \centering
    \includegraphics[width=\linewidth]{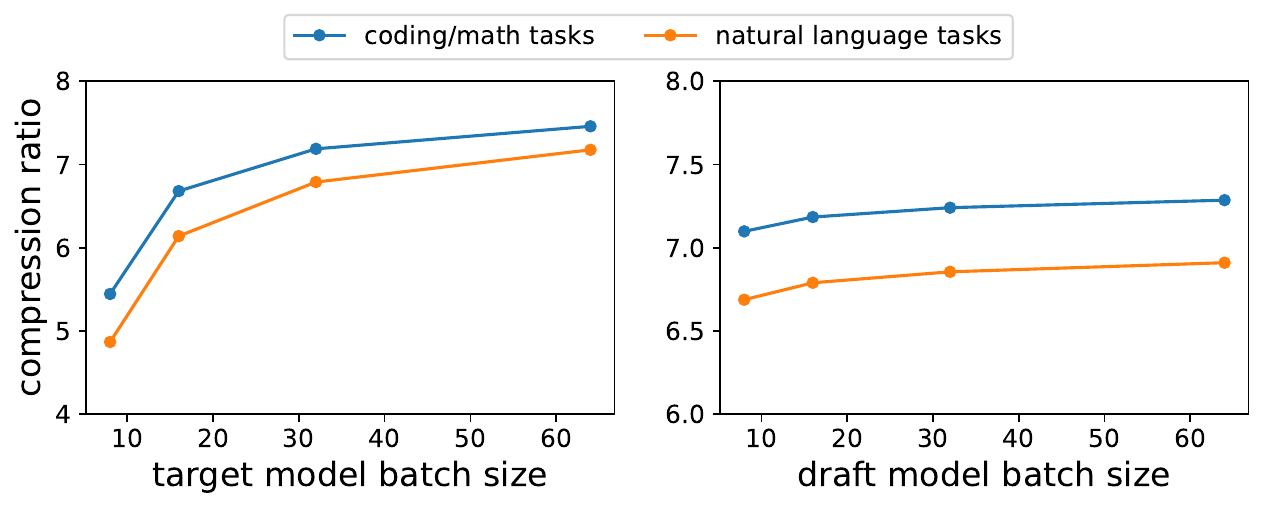}
    \vspace{-0.3in}
    \centering
    \caption{Compression ratio (average number of correctly guessed tokens per target model inference) while increasing the batch size of the target model (left figure) and draft model (right figure) }
    \label{fig:motivation_batchsize}
    % \size{what is compression ratio? is it hit rate?}
    % \label{fig:motivation_batchsize} \zy{it is the average number of accepted tokens per target model batch inference}
    % \vspace{-0.2in}
    % \vspace{-15pt}
    \vspace{-0.15in}
\end{figure}

\subsubsection{Low Batch Sizes in Tree-based Speculative Decoding} Figure~\ref{fig:motivation_batchsize} shows the effects of increasing the batch size for the target and draft models when serving a single request.  
We use Llama3 70B as the target model and Llama3 3B as the draft model, and we generate 7 tree layers (i.e., running the draft model 7 times) before running a batch (size 7) inference using the target model. In the first experiment (the left figure), we fix the batch size of the draft model to be 16, and the average number of guesses only increases marginally after the batch size is larger than 16. In the second experiment (the right figure), we fix the batch size of the target model to be 32, and when we increase the batch size of the draft model over 8, the average number of correct guesses of the target model almost stops increasing.

As we can see, to minimize the latency of serving a single request, it is usually sufficient to use a small batch size ($\leq 16$) for both the target and draft models. Note that this characteristic is different from throughput-oriented systems \cite{orca, distserve, sarathichunkedprefill}, where larger batch sizes improve system throughput while satisfying SLOs, rather than minimizing the end-to-end generation of a single request, which is the focus of our work.

\subsection{Key Limitations of Prior Work}

% \zy{Motivates that in single-request serving, both target model and draft model need smaller batch size }

This section describes how existing speculative decoding systems fall short in addressing the following key system challenges in single-request serving: (1) independent scalability across draft and target models, (2) KV cache consistency under speculative execution, and (3) kernel under utilization under low batch sizes. 

Table~\ref{tab:motivation_baseline_comparison} summarizes the differences between recent work, and the following paragraphs provide a detailed analysis on why they fail to address those three challenges.

\textbf{Lack of scalability of the target and draft models under low batch size.} For sequential speculative decoding approaches, since there is a strict dependency between the target and draft models, tree-based speculative decoding can only run one draft model and the target model at the same time. Therefore, to use parallelism for speculative decoding, we have to scale both the draft and target models across all GPUs; otherwise, some GPUs will be idle when running the models. However, using more GPUs for TP does not necessarily decrease latency. Table~\ref{tab:infer_time_intro} shows that the benefit of using more GPUs diminishes when we increase the degree of tensor parallelism for the Llama model family. It also shows that different models could have different compute requirements. For example, the 70B model benefits from using 4 GPUs instead of 2, while there is no benefit from using more than 2 GPUs for the smaller models (1B, 3B, 8B). Prior work using sequential speculative decoding (SpecInfer\cite{Miao_2024}, EAGLE \cite{li2024eagle2fasterinferencelanguage}, etc) fails to address this lack of scalability, while our design of parallel tree generation does (\S\ref{sec:parellel_tree_generation}).

\textbf{Lack of effective KV Cache management scheme for parallel tree-based speculative decoding.} Tree-based speculative decoding imposes a challenge on consistent and effective KV cache management. While parallel sequence-based speculative decoding (PEARL\cite{liu2024parallelspeculativedecodingadaptive}, AMUSD\cite{ mcdanel2024amusdasynchronousmultidevicespeculative}) has a concise way to manage the KV cache, only generating a chain of draft tokens yields a suboptimal compression ratio and thus low end-to-end performance. It is non-trivial to keep the consistency between the KV cache of both models and the draft trees. PipeInfer \cite{butler2024pipeinferacceleratingllminference} runs draft tree generation and target model verification in parallel and generates multiple draft trees to fill up the bubbles during execution. However, if some draft trees get invalidated, every draft tree generated after that will be discarded, wasting the compute and yielding low effective utilization. Therefore, to achieve a high performance, it is important---but challenging---to keep the KV cache consistent while maximizing the reuse of generated tokens. Our consistency management scheme addresses this (\S\ref{sec:kv_cache_management}).

% \zy{Low utilization of the kernels under low batch siz: all reduce operation have low-uiltization of the bandwidth, gemm/attention have low-utilization of the GPU}
\textbf{Low utilization of the GPU kernels.} While the computation and communication kernels are usually well-optimized for large batch inference, with large computation and communication demands, they do not perform well under low batch size.

 % \hank{IS this HBM or NVLink?  I think it should be specified}
Table~\ref{tab:utilization} shows the amount of bandwidth and compute utilization of the most time-consuming kernels in a transformer layer of a LLama 70B model split across 4 GPUs under a batch size of 8. The bandwidth utilization shown for the all-reduce operation is the NVLink bandwidth, while for the other operators, it is the HBM to SRAM bandwidth utilization. Under this low batch size, all operators are communication-intensive, yet the bandwidth utilization (both across GPUs and within a GPU) is low ($<10\%$) for both the all-reduce and attention operators. This is because the amount of communication is small, and therefore, the time is mainly spent on synchronization and waiting for the first input to arrive in the GPU SRAM. Some frameworks (vLLM \cite{vllmkwon2023efficientmemorymanagementlarge}, SGLang \cite{zheng2024sglangefficientexecutionstructured}, etc) use the NCCL LL protocol to reduce the latency in the all reduce operator when communication volume is low. However, as shown in Table \ref{tab:motivation_baseline_comparison}, there is still a significant opportunity to improve kernels to reduce latency for small batch sizes. Our latency-optimized kernels (in \S\ref{sec:low_latency_kernels}) fuse multiple optimizations to realize this opportunity.

\begin{table}[]
\resizebox{\columnwidth}{!}{%
\begin{tabular}{r|rrr}
Operators       & Time (in us) & Comp. Util. (\%)      & Band. Util. (\%)                         \\ \hline
QKV projection  & 16.9         & 2.0\%             & 18.7\%                                 \\
mask attention  & 18.8         & \textless{}0.01\% & \cellcolor[HTML]{FFCCC9}\textbf{6.5\%} \\
O projection    & 10.8         & 2.5\%             & 23.3\%                                 \\
all reduce & 12.0 & \textless{}0.01\% & \cellcolor[HTML]{FFCCC9}{\color[HTML]{333333} \textbf{8.5\%}} \\
SwiGLU          & 39.3         & 4.8\%             & 44.6\%                                 \\
down projection & 18.1         & 5.2\%             & 48.5\%                                 \\
all reduce      & 15.3        & \textless{}0.01\% & \cellcolor[HTML]{FFCCC9}\textbf{6.6\%}
\end{tabular}%
}
\caption{The utilization of the most time-consuming kernels in an int4 AWQ Llama 70B model when running TP on 4 GPUs.}
\vspace{-0.15in}
\label{tab:utilization}
\end{table}

% \section{the case of scaling single request serving}

% \input{sections/overview}

\section{\paper{} System Design and Architecture}
\label{sec:spec_pipe}

\paper{} addresses the three key systems challenges identified in the prior section through a modular design built around: (1) parallel tree generation, which enables asynchronous decoding and independent GPU allocation (\S\ref{sec:parellel_tree_generation}); (2) KV-cache consistency management, which supports reuse and correctness under speculative execution (\S\ref{sec:kv_cache_management}); and (3) latency-optimized fused kernels, which reduce communication and compute overhead under tensor parallelism (\S\ref{sec:low_latency_kernels}). We describe each component in turn.

\subsection{Parallel Tree Generation}
\label{sec:parellel_tree_generation}

To enable independent scaling of draft and target models, we introduce parallel tree generation, which splits the decoding process across GPU groups separately dedicated to drafting and verification. This allows both models to operate concurrently and avoids placing the draft phase on the critical path. The two groups communicate using NVLink/cross-network interconnect. The draft tree GPUs manage the draft tree and run draft inference to generate new tree nodes, while the target GPUs run the target model. Both the draft and target models are split across their respective GPUs through tensor parallelism (TP). GPUs computing the same model are connected tightly using NVLink.

Algorithm~\ref{alg:paralell_tree_algo} details the interaction between draft and target models in each decoding iteration. Denote one round (or one iteration) as the procedures of the draft and target model between two synchronization points.  Define $bs$ as the batch size of the target model, $w$ as the number of leaves for which we run the draft model inference (and thus expand the leaves to get potential children) each round (i.e., the batch size of the draft model), $d$ as the number of tree expansions in one round. Both target worker and draft worker run in a loop until the end of the generation, and synchronize when each finishes one iteration in the loop. 

In one iteration, the draft worker expands the draft tree $d$ times, by running inference on $w$ unexpanded leaves (i.e., the leaves where the KV-cache and logits (probability distribution of the next token) are not yet calculated) from the tree with the highest probability. After that, it synchronizes with the target worker to get the verified tokens. Then it re-roots the draft tree by walking down the tree using the path representing the verified tokens and adjusts the KV cache to stay consistent. After that, it grows the draft tree if there are not enough nodes to send to the target, and then it sends a sub-graph of the draft tree of size $bs$ to the target worker.

On the other hand, in one iteration, the target model constantly gets the draft tokens from the draft tree and runs batch inferences to calculate the logits. After that, it samples through the logits to generate the tokens one by one and then sends the verified tokens back to the draft worker.

\begin{figure}
    \centering
    % \vspace{-0.10in}

    \includegraphics[width=\linewidth]{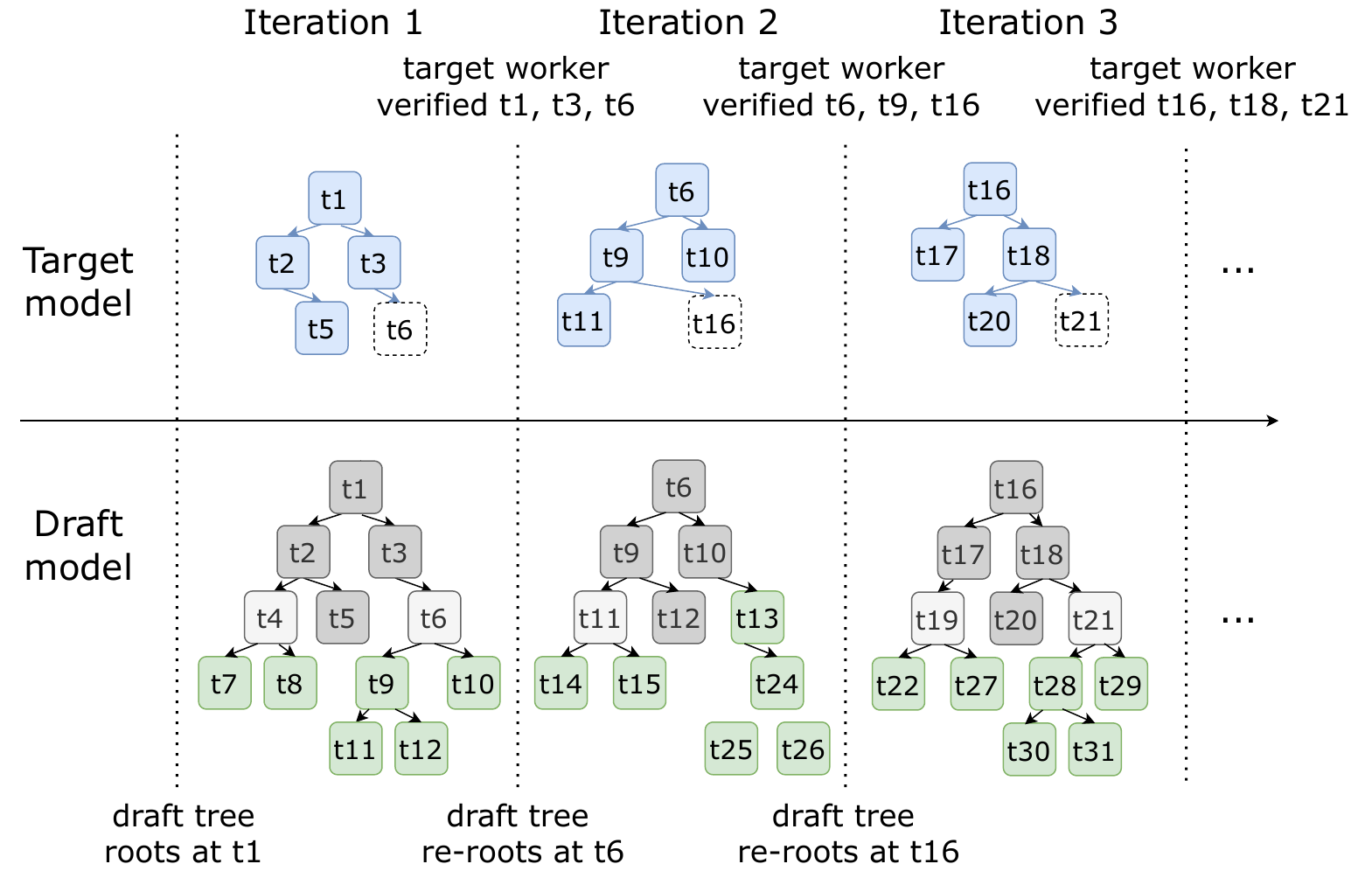}
    \caption{Example of parallel tree generation.  Each iteration re-roots the tree based on verified tokens, while the KV-cache is reorganized to preserve useful state.}

    \label{fig:execution}
\end{figure}

Figure~\ref{fig:execution}  shows three decoding iterations. In each iteration, the draft model grows the tree while the target model verifies a subgraph. The tree is then re-rooted, and verified tokens are promoted to the KV cache. In this example, $bs = 4$, $d = 3$, $w = 2$. 

At the start, the draft tree is \(t_1, t_2, t_3, t_4, t_5, t_6\), and the draft workers select the top \(bs=4\) tokens \((t_1,t_2,t_3,t_5)\) to give as \(input_1\) to the target workers. During iteration 1, while the draft workers continue growing the tree with 6 new nodes, the target workers run inference on \(input_1\) and sample \(output_1=(t_1,t_3,t_6)\). Then, the draft workers verify that \((t_1,t_3,t_6)\) is a valid path in the tree and re-root at \(t_6\). With enough nodes remaining, they choose the next top 4 tokens \((t_6,t_9,t_{10},t_{11})\) as \(input_2\). During iteration 2, the draft workers grow 6 more nodes while the target workers process \(input_2\) and produce \(output_2=(t_6,t_9,t_{16})\). However, \(t_{16}\) is not yet in the tree, so the draft workers re-root at \(t_{16}\) and keep growing new nodes \(t_{17}, t_{18}, t_{19}, t_{20}, t_{21}\), giving \((t_{16}, t_{17}, t_{18}, t_{20})\) as \(input_3\). During iteration 3, a similar process continues, with the draft and target workers running in parallel, growing and verifying tokens as they build out the tree.

\textbf{Maximum-likelihood tree expansion} We use the logarithm of the softmax probability as the value of each node, and use the sum of values from the root to each node as the weight. Thus, a higher weight means a higher probability that a token could be generated (under the distribution of the draft model). We keep the pair $(value, node)$ in a priority queue to efficiently get the most probable leaves in $O(k \log{s})$, where $s$ is the number of probable leaves to consider and expand the tree.

\begin{algorithm}[t]

\SetAlgoLined 
\small
% \KwIn{}
\KwIn{Depth of tree to generate $d$, width of tree to generate $w$, batch size of target model $bs$}

% \SetKwInOut{Input}{input}
% \Input{}\
\If{Is\_draft\_worker} {
    \While{True} {
    
        \For{$i = 1$ to $d$}{
            Expand the $w$ most probable leaves from the tree\; 
        }
        Get verified tokens from the target worker\;
        \If{target worker signal to stop } {
            break\;
        }
        Update KV Cache and draft tree based on verified tokens\; 

        \While{Tree size $ < bs$} {
            Expand the $w$ most probable leaves from the tree\;
        }
        Get the most probable subgraph of size $bs$ from the draft tree\;
        Send it to the target worker to verify\;
    }
}\ElseIf{Is\_target\_worker} {
    \While{True} {
        Get draft tokens from the draft tree worker\;
        Verify the draft tokens by batch inference\;
        \If{Reach the end of generation}{
            Send the stop signal to the draft worker\;
            Break\;
        } \Else{
            Send the verified tokens to the draft tree worker\;
        }
    }
}

\caption{Parallel tree generation algorithm} 
\label{alg:paralell_tree_algo}
\end{algorithm}

\textbf{GPU allocation for draft model and target model} Given a GPU node of $k$ GPUs, we will allocate $x (1 \leq x \leq k - 1)$ GPUs to the target model and $(k - x)$ GPUs to the draft model. To determine which $x$ to use, we run a profile phase before serving the queries, where we try out different $x$s to find which configuration yields the fastest average decoding speed. We found that if we fix the target model, the optimal $x$ is smaller when we are using a more powerful draft model (more analysis is in \ref{sec:eval_design_choices})

\textbf{Setting the batch size} Larger $bs, w$ will lead to higher acceptance ratio per iteration, but when $bs, w$ get larger and larger, the margin gain on the acceptance ratio will decrease, and total running time will increase. We set $bs = 8$ and $w = 8$ empirically to balance the acceptance ratio and running time, based on our analysis in \S$\ref{sec:eval_design_choices}$.

\begin{figure}[t]
    \centering
    \includegraphics[width=0.9\linewidth]{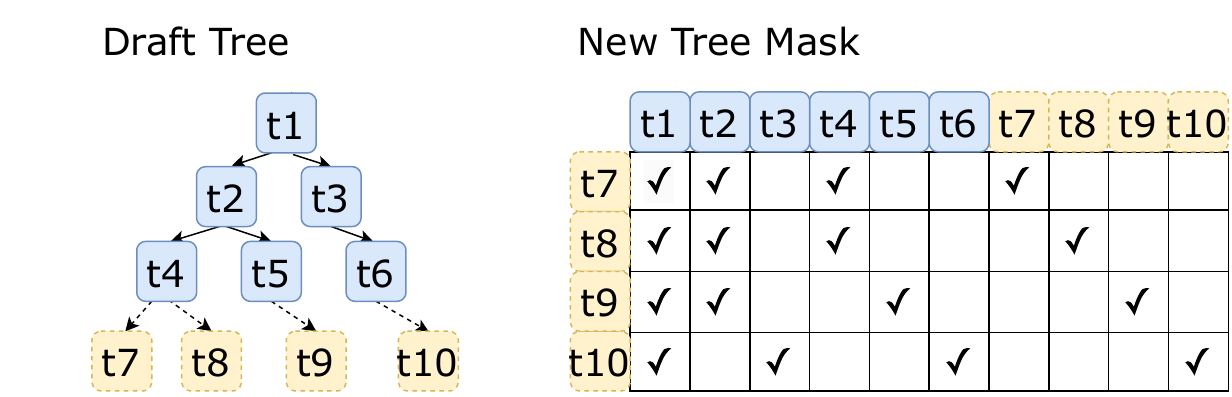}
    \caption{Example of a non-square tree mask during draft tree expansion: yellow nodes are the leaves to expand, and the blue nodes are the existing tree nodes}
    \label{fig:tree_mask}
\end{figure}
\textbf{Setting the number of tree expansions $d$ in one round.} 
Before we start serving the requests, we first profile the running time of both the draft model and the target model. Denote $t_{target}$ as one round of target model inference, and $t_{draft}$ as one round of draft tree expansion. Define $r = \lfloor \frac{t_{target}}{t_{draft}} \rfloor$. We set $d = r$ or $d = r + 1$ so that draft tree expansion and the target model verification finish nearly at the same time to maximize parallelism.

\textbf{Non-square mask support for efficient masked attention kernel} The attention operator in the target model uses a square mask, since the target model takes a tree each time, and each token will only mask out the attention with those tokens that are not the ancestors within the current input. This is similar to prior work \cite{Miao_2024}. However, for the draft model, this is not the case. Consider the example in Figure$~\ref{fig:tree_mask}$ with a current tree of size $6$, and we want to calculate the logits of $4$ probable leaves, then regarding the tree cache, we only calculate the attention of each leave with its ancestor on the tree (and also all the data that is in the prefix cache). In this case, we need a mask of at least size $(4, 10)$ to contain all the necessary information. Therefore, we support a non-square mask as input in our attention operator for the draft model.

\begin{figure}
    \centering
    % \vspace{-0.10in}
    \includegraphics[width=\linewidth]{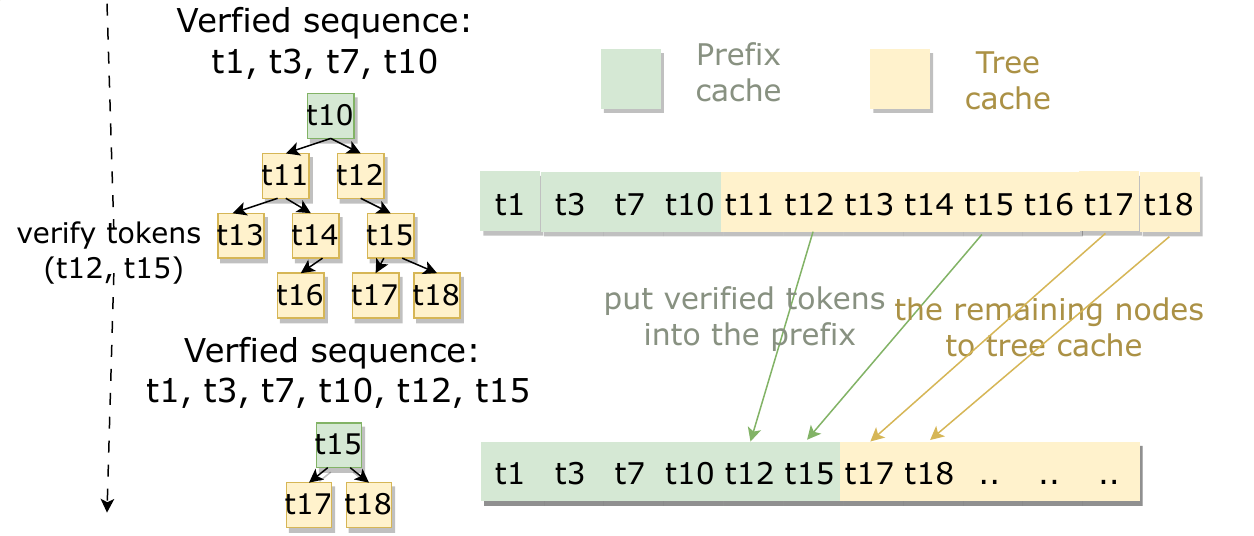}
    \caption{KV cache management of the draft model: each time after the new verified tokens get updated, the KV states of the verified tokens will be in the prefix of the KV cache (green part), and the KV states of the draft tree tokens will be right after the prefix (yellow part).}
    \label{fig:kv_cache_upd}
\end{figure}

\subsection{KV Cache Consistency Management} 
\label{sec:kv_cache_management}

To maintain the consistency between the draft model and target model, we develop a consistency management scheme to reorganize the KV cache of the draft model so that it remains consistent with the target model and the draft tree and maximizes the re-use of previously computed KV states.

Throughout the execution, we organize the KV Cache of the draft model as follows and keep it as an invariant throughout the execution: the KV states of the verified tokens are stored continuously in the prefix of the KV cache (which we call prefix cache), and the KV states of the tree are stored right after the prefix (which we call tree cache).

While the target worker is doing batch inference and sampling on the tokens from the draft workers, the draft worker keeps generating new tree leaves in the draft tree, appending the tree cache after the existing entries.

\textbf{Re-organization of KV cache for verified tokens} After the target worker samples the tokens, it sends the verified tokens to the draft worker. The draft worker then walks through the tree using the verified tokens and reroots at the last verified token. Then, if the last verified token exists in the current draft tree, then we will reorganize the tree cache so that only the KV states of the nodes in the new subtree remain in the tree cache. In this way, even when some of the predicted tokens we send to the target worker are wrong, we can still reuse all the computed KV states in the subtree, avoiding any recomputation.

Figure~\ref{fig:kv_cache_upd} shows an example of how the KV cache of the draft model is updated when there are new verified tokens. Suppose the sequence $(t_1, t_3, t_7, t_{10})$ is already verified, and the prefix of the cache is the KV states of those tokens, and the KV states of the draft tree tokens are organized contiguously after the prefix. When we update the verified tokens to be $(t_1, t_3, t_7, t_{10}, t_{12}, t_{15})$, we walk down the draft tree using the newly verified tokens $(t_{12}, t_{15})$. Then we reach the node $t_{15}$, which means the nodes in the subtree, $t_{17}, t_{18}$, are still useful in the future.  Therefore, we move $t_{12}, t_{15}$ to the prefix cache so that it stores the information of the same verified tokens as the target model. Then it reorganizes the remaining sub-tree of $t_{15}$ (i.e. $t_{17}, t_{18}$) into the next positions available, discarding the KV states that are no longer useful (e.g. $t_{11}, t_{12}$, etc).

In the case where the draft tree does not have enough nodes to send back to the target worker, it expands $bs$ nodes immediately using one draft model inference. In either case, the draft tree will have enough nodes to pass to target workers, therefore entering the next iteration, with the KV states synchronized across the draft model, the target model, and the draft tree.

\subsection{Latency-Optimized Kernels Design}
\label{sec:low_latency_kernels}
To reduce the inference time of both draft and target under low batch size, we design and implement  \textit{latency-optimized} operators for the LLama and Qwen model family. While our design could be applied to any precision, we implement it for the int4 AWQ quantized model. The operators we optimize include all-reduce, masked attention, and SwiGLU. We first introduce the Nvidia Collective Communication Library's Low Latency (NCCL LL) protocol, which our work leverages heavily.

\begin{algorithm}[t]
\SetAlgoLined
\SetKwFunction{FuncStoreLL}{storeLL}
\SetKwProg{Fn}{Function}{:}{}
\Fn{\FuncStoreLL{$\text{dst: address}$, $\text{val: 64-bit int}$, $\text{flag: 32-bit int}$}}{
    $val\_low \gets \text{lower 32 bits of } val$\;
    $val\_high \gets \text{upper 32 bits of } val$\;
    AtomicStore($dst$, $val\_low$, $flag$, $val\_high$, $flag$)\;
    % Store values global memory using the following assembly command:\;
    % \texttt{"st.volatile.global.v4.u32 [dst], \{val\_low, flag, val\_high, flag\};"}\;
}
\SetKwFunction{FuncReadLL}{readLL}
\SetKwProg{Fn}{Function}{:}{}
\Fn{\FuncReadLL{$\text{src: address}$, $\text{flag: 32-bit integer}$}}{
    % $src \gets \text{recvPtr}(i) + \text{offset}$\;\\
    % $flag \gets \text{recvFlag}(i)$\;\\
    $data1, flag1, data2, flag2 \gets 0$\;
    \Repeat{}{
        $flag1$, $data1$, $flag2$, $data2$ = AtomicLoad($src$) \;
    }{\texttt{flag1 = flag \textbf{and} flag2 = flag}}\;
    $val64 \gets data1 + \big((\text{uint64\_t})data2 \ll 32\big)$\;
    \Return $val64$\;
}
\caption{\texttt{storeLL} and \texttt{readLL} Function}
\label{alg:ll_protocol}
\end{algorithm}

\textbf{NCCL LL protocol.} This is a communication scheme to reduce the latency of GPU-to-GPU send and receive operations. Algorithm~\ref{alg:ll_protocol} shows the store and load function of the LL protocol.  The functions $\textit{AtomicStore}$ and $\textit{AtomicLoad}$ are NVidia GPU PTX instructions $\textit{ld.volatile.global.v4.u32}$ and $\textit{st.volatile.global.v4.u32}$, which load and store 16 bytes of data from and to the global memory in the same or different GPUs. Each 8-byte (2 32-bit unsigned integers) load/store is atomic.

The storeLL function takes a 64-bit integer $val$ and a 32-bit integer flag, splits $val$ into two 32-bit integers, and stores them each with $flag$ to a memory location. The loadLL function takes a memory location and a flag. It keeps polling until the flag at the memory location matches the expected one, and then combines the 2 32-bit integers into a 64-bit integer to return.

Using those two functions, we can implement a communication scheme without any explicit synchronization. Assume that last time we store some value $x$ as a flag, and this time we use $x + 1$. 
The other compute unit (another stream multi-processor in the same GPU or on another GPU) will know the data is ready when it sees $x + 1$ as the flag, without any additional synchronization. In our latency-optimized kernels, we heavily rely on this powerful primitive to reduce synchronization overhead both within and across GPUs.

\begin{figure}
    \centering
    \includegraphics[width=0.9\linewidth]{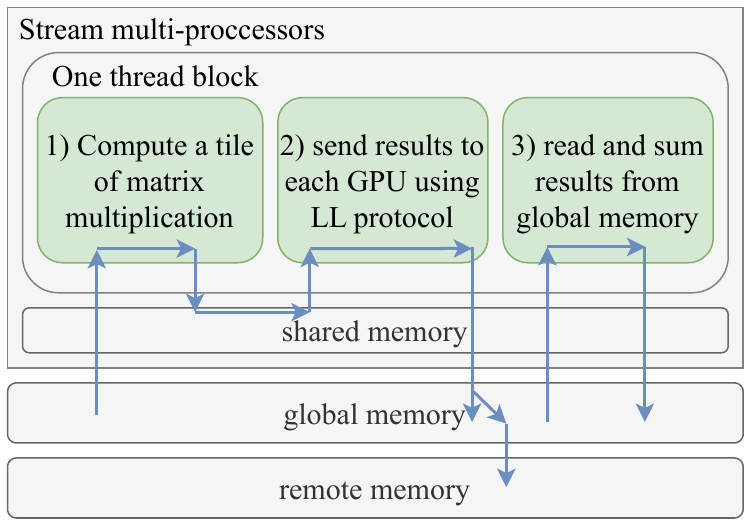}
    % \vspace{-0.15in}
    \centering
    \caption{Execution flow of fused GEMM all-reduce operator}
    \label{fig:fused_allreduce}
    % \vspace{-0.2in}
    % \vspace{-15pt}
\end{figure}

% In our all reduce operators, we adopt 
\textbf{Fused GEMM with all reduce.} 
To further reduce the data movement overhead and save the number of synchronization barriers, we fuse each all-reduce operation with the preceding GEMM operation. Figure~\ref{fig:fused_allreduce} shows the computation and data flow within one thread block when we run GEMM fused with all reduce. Each thread block has three steps during execution:
\begin{itemize}
    \item The threadblock calculates a contiguous set of columns in the final result matrix.
    \item The threadblock starts sending the result to each GPU using the NCCL LL protocol.
    \item The threadblock on each GPU waits until the results are available, aggregates them to get the final results, and stores them back to the global memory on the current GPU.
\end{itemize}
The synchronization between steps 2 and 3 is implemented using just the NCCL LL protocol described above and no additional, expensive mechanisms.

\textbf{Masked attention.} For the mask-attention operators, we fuse the position embedding with the attention calculation. Then, within one GPU, we first split the computation of the single attention head between different thread blocks. After the calculation, the threadblocks aggregate the sum using the NCCL LL protocol within a single GPU. Similar to our Fused GEMM with all-reduce operators, we add up the results within one attention head without explicit synchronization across thread blocks or extra kernel launches.

\textbf{Fused SwiGLU.} This operator is of the form $SwiGLU(x,$  $W, V, b, c) = \sigma (xW + b) \oplus (xV + c)$. We implement tile-based matrix multiplication, where each threadblock calculates the same tile of the two matrix multiplications. This avoids loading the input twice from the GPU HBM. 
Right after we get the output of the tiles, we calculate the sigmoid and dot product before putting the results back to the GPU memory, avoiding unnecessary data movement.

\section{Implementation}
\label{sec:implementation}
We implement our latency-optimized kernels using $\sim 3000$ lines of CUDA and C++, and our tree-based parallel speculative decoding using $\sim 4000$ lines of C++ and Python. We use the CUTLASS library \cite{Thakkar_CUTLASS_2023} to implement the fused, latency-optimized kernels.  

Aside from the optimized, fused all-reduce operators within the decoder layers in the transformer models, we use the NCCL library to communicate and synchronize states between different workers.

We wrap our optimized CUDA kernels in Python and capture all the kernels in a forward pass into a CUDAGraph for both our draft model and target model to reduce kernel launch overhead. 

% \textbf{Bookkeeping draft tree} In each node in draft tree, we keep it depth, weight, and the list of ancestor nodes, so that we could calculate the mask quickly using the list of ancestor nodes, the get the position id for positional embedding using the depth, and calculate the 

% Please add the following required packages to your document preamble:
% \usepackage{graphicx}
\begin{table}[t]
\resizebox{\columnwidth}{!}{%
\begin{tabular}{c|c}
Target model                  & Draft model                      \\ \hline
Llama-3-70b-Instruct          & Llama-3.2-3B, EAGLE-0.99B        \\
deepseek-coder-33b-instruct   & deepseek-coder-1.3b-instruct     \\
Qwen2-72B-Instruct            & Qwen2-1.5B-Instruct, EAGLE-1.05B \\
DeepSeek-R1-Distill-Qwen-32B  & DeepSeek-R1-Distill-Qwen-1.5B    \\
DeepSeek-R1-Distill-Llama-70B & DeepSeek-R1-Distill-Llama-8B    
\end{tabular}%
}
\caption{The set of models we use in our evaluation of our approach and baselines }
% \vspace{-0.3in}
\label{tab:models_eval}
\end{table}

\textbf{Saving the number of CUDAGraphs for variable-sized masks.} A CUDAGraph only supports a fixed-shape input. To reduce the number of CUDAGraphs that we need to capture, we perform a small optimization.  We pad all tree masks to a fixed-size shape of ($w$, $maxseqlen$), where $w$ is the number of leaves to expand, and we copy the tree mask of size $(w, w+T)$, where $T$ is the size of the tree, into the suffix of this larger structure and mask the prefix to be all $1$'s.  This allows us to reuse CUDAGraphs with variable-sized inputs.  Specifically, we only need to capture $O(w_{max})$ (which is less than 20) CUDAGraphs, while supporting tree masks of more than one thousand different shapes. 

\textbf{Enabling arbitrary tensor parallelism for transformer models.} Llama models (and transformer models from other families) usually need to satisfy two constraints before running tensor parallelism across $x$ GPUs: (1) the dimension of the matrix multiplication (after split across $x$ GPUs) is divisible by the GEMM kernel's block size  (usually 128 or 256); and (2) the number of attention heads must be divisible by $x$ so that each attention head can be distributed to a single GPU. When we split the LLama 70B model across a power-of-two number of GPUs (i.e, 1/2/4/8), those two requirements are naturally satisfied since the number of attention heads and the dimension of the 70B model are divisible by a large power of 2. 

However, when we want to enable a more fine-grained allocation of the draft model and the target model (e.g., using 6 GPUs for the target model and 2 GPUs for the draft model), those two requirements may not be satisfied. To resolve this, we increase the dimension of the matrix operation so that they are divisible by the number of GPUs $x$ and zero-pad. For the attention calculation, we increase the number of attention heads so that it is divisible by $x$, and we also pad the projection matrix before and after the multi-head attention so that the calculation of padded attention heads does not contribute to the final result and the model output is equivalent of the non-padded model.

% \zy{ read the code and see if there is things left to cover (like accelerate using prioirty queue)}
\section{Evaluation}

% In our evaluation, we want to answer and analyze the following questions:
% \begin{itemize}
%     \item How large is the improvement using our low-latency kernels, regarding single operators and model end-to-end inference time?
%     \item How much improvement we have using our parallel tree-based speculative decoding over other speculative decoding paradigms?
%     \item How much improvement do we have combining both techniques over the existing LLM serving systems?

% \end{itemize}

% % $\~ref{}$ will focus on 
%  We compare the performance of \paper{} against three main-stream LLM serving systems, vLLM \cite{vllmkwon2023efficientmemorymanagementlarge}, SGLang \cite{zheng2024sglangefficientexecutionstructured} and TRTLLM \cite{trtllm}

% Our evaluation 
% The datasets used were MT-bench (Zheng et al., 2023), HumanEval (Chen et al., 2021), GSM8K (Cobbe et al., 2021),
% Alpaca (Taori et al., 2023), CNN/Daily Mail (Nallapati et al.,
% 2016), and Natural Questions (Kwiatkowski et al., 2019).

In our evaluation, we want to answer and analyze the following questions:
\begin{itemize}
    \item What is the end-to-end improvement of \paper{} compared to other speculative decoding baseline systems? (in \S\ref{sec:eval_e2e})
    \item How much does parallel tree generation (with kv-cache synchronization in asynchronous decoding) improve the end-to-end performance? (in \S\ref{sec:eval_alabation})
    \item How much does latency-optimized kernels improve the end-to-end performance? (in \S\ref{sec:eval_alabation})
    \item What is the improvement of single operators using our latency-optimized kernels? (in \S\ref{sec:eval_ll_kernels})
    \item What are the design choices we make to achieve good performance using our framework? (in \S\ref{sec:eval_design_choices})

\end{itemize}

% Our evaluation in \ref{sec:eval_e2e} shows that \paper{} could serve these two models with a consistently higher decoding speed than the other two baseline systems. We then perform a deeper analysis to break down our performance gain from our two techniques: disaggregated tree generation (\ref{sec:eval_spec}) and low-latency kernels (\ref{sec:eval_ll_kernels}). After that, we look closer into different design choices (including batch size in speculative decoding, GPU allocation, etc) in our ablation studies section (\ref{sec:eval_ablation}).

\subsection{Experiments setup}

\noindent \textbf{Cluster setup} We evaluate \paper{} and the baselines on one node with 8 NVIDIA
80GB H800 SXM GPUs connected by NVLink, which is the most common setup of a Nvidia Hopper compute node. We use all 8 GPUs to minimize the decoding latency for all of our experiments, except that, in \S\ref{sec:eval_ll_kernels}, we show the performance improvement of our latency-optimized kernels under a subset of these GPUs.

\noindent  \textbf{Models and model configurations} We evaluate our system under five different pairs of models (as shown in \ref{tab:models_eval}) from different families, including LLama3 \cite{touvron2023llamaopenefficientfoundation},  Deepseek-Coder \cite{guo2024deepseekcoderlargelanguagemodel}, Qwen2 \cite{yang2024qwen2technicalreport}, Deepseek R1-Distilled Qwen, and Deepseek R1-Distilled Llama \cite{deepseekai2025deepseekr1incentivizingreasoningcapability}. Deepseek-Coder is a series of models that focus on coding, while the rest of the model families have general abilities. 
From each model family, we pick a large model as the target model (generally having $>30$B parameters) and a small model as the draft model ($<10$B). For Llama3 and Qwen2 family, there are also trained EAGLE2 models, so we used those in the baselines, which supports EAGLE-based \cite{li2024eagle2fasterinferencelanguage} speculative decoding. We apply 4-bit AWQ quantization with a group size of 128 \cite{lin2024awqactivationawareweightquantization} to all the weights of the transformer layers models in each family except the EAGLE models. We keep the BF16 precision for the embedding layers and the LM head operator. Each model uses BF16 to compute the attention and the linear operators (after weight de-quantization). We apply the same quantization to both the baseline and our system, and therefore, even with the latency-optimized kernels, the computation of each single model in our system is equivalent to that of each baseline model.

% $\cite{llama}$

\noindent \textbf{Datasets} We evaluated our system in six different datasets across different domains: MT-bench \cite{zheng2023judgingllmasajudgemtbenchchatbot}, HumanEval \cite{chen2021evaluatinglargelanguagemodelshumaneval}, GSM8K \cite{cobbe2021trainingverifierssolvemathgsm8k}, Alpaca \cite{alpaca}, CNN/Daily Mail \cite{nallapati2016abstractivetextsummarizationusingcnndaily}, and Natural Questions \cite{kwiatkowski-etal-2019-naturalquestions}. The datasets consist of tasks including summarization, math questions, coding questions, etc.

% Please add the following required packages to your document preamble:
% \usepackage{graphicx}
\begin{table}[t]
\resizebox{\columnwidth}{!}{%
\begin{tabular}{c|ccccc}
              & EAGLE        & smaller draft & draft TP     & tree spec   & performance \\ \hline
vLLM          & {\color[HTML]{009901}$\checkmark$} & {\color[HTML]{009901}$\checkmark$}  & {\color[HTML]{FE0000}$\times$}     & {\color[HTML]{FE0000}$\times$}   &  {\color[HTML]{FE0000}not optimal}    \\
SGLang        & {\color[HTML]{009901}$\checkmark$}    & {\color[HTML]{FE0000}$\times$}       & {\color[HTML]{009901}$\checkmark$} & {\color[HTML]{009901}$\checkmark$} &  {\color[HTML]{009901}optimal (EAGLE)} \\
TRT-LLM       & {\color[HTML]{009901}$\checkmark$}&{\color[HTML]{009901}$\checkmark$} & {\color[HTML]{009901}$\checkmark$} & {\color[HTML]{FE0000}$\times$}   &  {\color[HTML]{FE0000}not optimal}  \\
\paper-base & {\color[HTML]{FE0000}$\times$}  &{\color[HTML]{009901}$\checkmark$} & {\color[HTML]{009901}$\checkmark$} & {\color[HTML]{009901}$\checkmark$}  &  {\color[HTML]{009901}optimal (smaller draft)}
\end{tabular} %
}
\caption{The support matrix for different baselines in the evaluation under int4 AWQ}
% \vspace{-0.4in}
\label{tab:baseline_support}
\end{table}

 % \cellcolor[HTML]{FFFFFF}{\color[HTML]{009901} Tree} &
 %    \cellcolor[HTML]{FFFFFF}{\color[HTML]{009901} 

\noindent \textbf{Baseline systems} We compare our solution against three major LLM serving frameworks: vLLM, SGLang, and TensorRT-LLM.
\begin{itemize}
    \item vLLM \cite{vllmkwon2023efficientmemorymanagementlarge}: vLLM supports both EAGLE and a smaller model from the same model family as the draft model. However, it only supports sequence-based speculative decoding for both cases.
    \item SGLang \cite{zheng2024sglangefficientexecutionstructured}: SGLang supports using the EAGLE model as the draft model, with tree-based generation.  Out of the five model families we benchmark on, only two pairs (Llama 70B and Qwen 72B) have the corresponding EAGLE draft model, and therefore, we benchmark the speed of the normal auto-regressive generation for the other three model families. 
    \item TRT-LLM \cite{trtllm}: Under int4-awq precision, TRT-LLM only supports sequence-based speculative decoding with a smaller draft model. 
\end{itemize}

Since the normal tree-based speculative decoding with tensor parallelism is not supported on vLLM and TRT-LLM, to make a fair comparison with the existing techniques, we also implement:
\begin{itemize}
    \item \paper-base: This is an implementation with only our optimized custom attention kernels. It uses the one-shot all-reduce implementation in the PyTorch Distributed library, and a SwiGLU implementation using normal AWQ multiplication kernel and PyTorch implementation of $silu$ activation and dot product. It also performs sequential speculative decoding with the draft model and the target model both running on all GPUs. 
\end{itemize}

Table \ref{tab:baseline_support} shows the configuration of speculative decoding that each baseline supports. As shown in later benchmarks, the most competitive baselines are SGLang and \paper-base since they support serial tree-based speculative decoding for EAGLE and a smaller draft, respectively. For each baseline implementation of each model, we run them in an extensive set of configurations and choose the configuration that maximizes the average tokens per second across all the datasets.

\begin{figure}
    \vspace{0.08in}

    \centering
    \includegraphics[width=\linewidth]{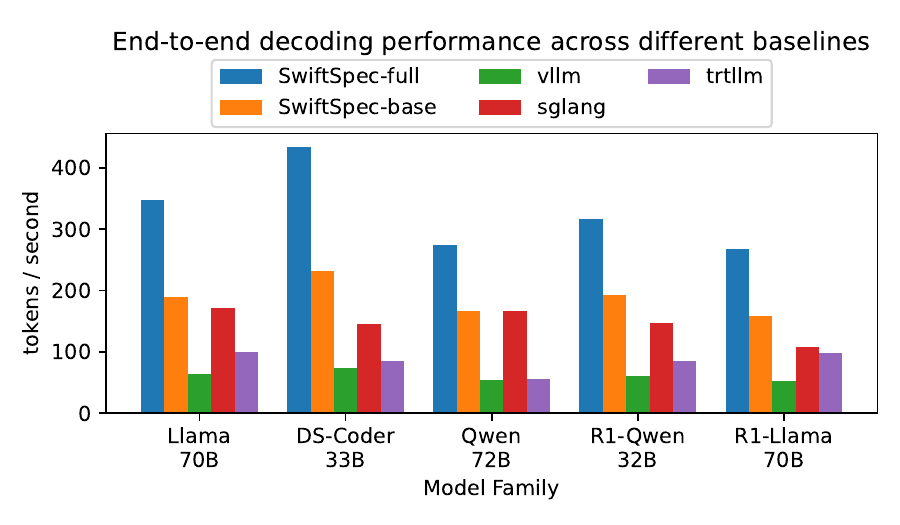}
    \vspace{-0.35in}
    \caption{End-to-end decoding speed comparison between baselines.}
    \vspace{-0.20in}
    \label{fig:perf_e2e}
\end{figure}

\subsection{End-to-end results}
\label{sec:eval_e2e}

Figure~\ref{fig:perf_e2e} shows the end-to-end benchmark on the decoding speed of \paper{} along with the four baselines, vLLM, SGLang, TRT-LLM, and \paper-base.  Due to the limitation that vLLM and TRT-LLM only support sequence-based speculative decoding, their performance is not comparable with SGlang and \paper-base. For the Llama 70B model and the Qwen 72B model, EAGLE speculative decoding is supported in SGLang, and thus SGLang has a comparable performance with \paper-based; for the other models (Deepseek-Coder, R1-Qwen, and R1-Llama), we benchmark SGLang with autoregressive decoding since EAGLE and other speculative decoding methods are not supported. Therefore, for those models, \paper-base serves as a baseline that we can compare more fairly. As shown in Figure~\ref{fig:perf_e2e},  our \paper{} system consistently outperforms \paper-base (by an average of $1.75\times$) and SGLang (on average $2.23\times$), the two most competitive baselines in the benchmark.

\begin{figure}
    \centering
    \includegraphics[width=\linewidth]{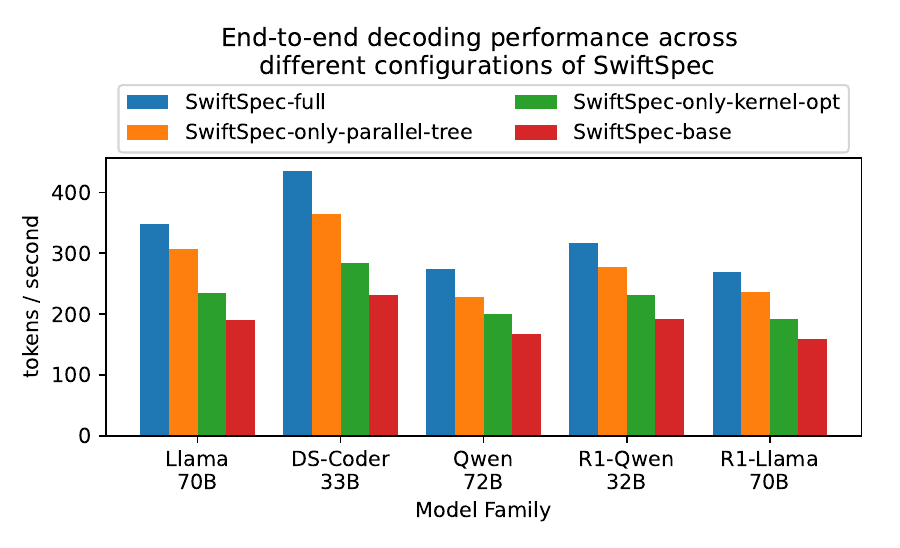}
    \vspace{-0.30in}
    \caption{End-to-end decoding speed comparison across
different configurations of \paper{}}
    \vspace{-0.15in}
    \label{fig:perf_ablation}
\end{figure}

\subsection{Ablation studies}
\label{sec:eval_alabation}

% \zy{compare falcon/falcon-only-kernel/falcon-only-parallel-spec/falcon-no-opt}

In this paper, we propose three techniques, which help two classes of optimizations: Parallel tree generation and consistency management together support efficient parallel tree-based speculative decoding, while latency-optimized kernels support more efficient kernel execution. We study the effect of each of those two classes of optimizations to reason about where our performance gain comes from. 

As shown in Figure~\ref{fig:perf_ablation}, we benchmark our framework under four different configurations of our framework: \circled{1} \paper, \circled{2} \paper-only-parallel-tree, \circled{3} \paper-only-kernel-opt, \circled{4} \paper-base. \paper{} and \paper-base are introduced earlier; \paper-only-parallel-tree is a version of \paper{} with only the parallel tree generation mechanism and the optimized attention kernels, while \paper-only-kernel-opt is a version of \paper{} with only the latency-optimized kernels (and serial speculative decoding).  Note that all four configurations contain our customized attention kernels since a non-square mask is needed to support our tree expansion algorithm, which is not supported by other works. We reason about the improvement of our attention kernel optimization alone later in this section and also in the kernel micro-benchmark section (\S\ref{sec:eval_ll_kernels}).

\textbf{Effect of parallel tree generation (with KV synchronization)}
To show the effect of our parallel tree generation with KV cache consistency management, we compare \circled{1} with \circled{3} (using vs not using parallel tree generation under latency-optimized kernels) and \circled{2} with \circled{4} (using vs not using parallel tree generation without latency-optimized kernels).  As shown in Figure~\ref{fig:perf_ablation}, \circled{1} outperforms \circled{3} by an average speedup of $1.43 \times$, and \circled{2} outperforms \circled{4} by an average speed of $1.50\times$.

\begin{table}[]
\resizebox{\columnwidth}{!}{%
\begin{tabular}{c|ccc|cccccc}
\hline
\multirow{2}{*}{} &  &\multicolumn{2}{c|}{model time} & \multicolumn{6}{c}{compression ratio for different datasets} \\
         &  Decoding 
speed&target & draft & ALP & GSM & HE & MT & QA  & SUM  \\ \hline
parallel &         275 tokens/s&10.48ms&       3.25ms& 2.92   & 3.78  & 3.92      & 3.11      & 2.74 & 3.56 \\
serial   &         200 tokens/s&10.34ms&       3.73ms& 3.28   & 4.2   & 4.04      & 3.42      & 3.12 & 3.72 \\ \hline
\end{tabular}%

}
\vspace{0.1in}
\caption{The comparison of compression ratio, model inference time, and decoding speed under parallel and serial tree generation for model Qwen2-72B and draft model Qwen2-1.5B. ALP, GSM, etc are the abbreviations of the six datasets we use in the benchmark.}
\vspace{-0.20in}
\label{tab:ablation_compare}

\end{table}

To look deeper into how our parallel tree generation mechanism brings an improvement of an average of $43\%$ (from \circled{3} to \circled{1}), we look specifically into the Qwen2 model family as an example, where the optimal configurations (including tree depth $d$, tree width $w$, etc) are the same for using and not using parallel tree generation. While the serial version uses 8 GPUs for both models, the parallel version uses 2 GPUs for the draft model and 6 GPUs for the target model.

Table~\ref{tab:ablation_compare} shows the compression ratio (average number of successful guesses per target model verification) and the target and draft model inference time under parallel/serial speculative decoding (\circled{1} and \circled{3}). When we are running parallel tree generation instead of serial speculative decoding, the draft model does not know the result of the concurrent target model verification, and thus it generates some nodes that are not useful after the verification. Therefore, the compression ratio is smaller when we use parallel tree generation. However, due to our maximum-likelihood expansion of the draft tree and the maximized reuse of the KV cache, the compression ratio is only $9\%$ less than the serial version on average. Furthermore, our draft model inference time decreases from $3.72ms$ to $3.25ms$ using only 2 GPUS instead of 8 GPUs saving $79\%$ of the GPU cycles during one round of drafting, and our target model has nearly no slow down using 6 GPUs instead of 8 GPUs (from $10.34$ms to $10.48$ms), still saving $25\%$ of the GPUs cycles during one round of target verification. As a result, the end-to-end tokens-per-second of Qwen2-72B model increases from 200 to 275 (by $37\%$) by applying our parallel tree generation and KV cache management scheme.

\textbf{Effect of latency-optimized kernels}
To illustrate effect of our latency-optimized fused GEMM-all-reduce kernel and SwiGLU operators, we compare \circled{1} with \circled{2} (using vs not using those kernels under parallel tree generation) and \circled{3} with \circled{4} (using vs not using those kernels under serial tree generation. ). As shown in Figure~\ref{fig:perf_ablation}, \circled{1} outperforms \circled{2} by an average speedup of $1.16 \times$, and \circled{3} outperforms \circled{4} by an average of $1.21\times$. Therefore, those two kernel optimizations give the end-to-end improvement of at least $16\%$ for both parallel and serial tree generation.  

To illustrate the improvement of our optimized attention kernel, we benchmark the Llama 3B model under TP=4 as an example. Under an input token length of 8 and a context length of 500, an inference of the 3B model with all the latency-optimized kernel optimizations takes 3.25 ms to finish, while an inference with a FlashAttention kernel takes 3.54 ms. This shows an $8.2\%$ reduction in the inference time of the Llama 3B model using our optimized attention kernel.  

Refer to  \S\ref{sec:eval_ll_kernels} for the inference time of individual operators to understand our per-operator improvement.

% Figure~\ref{fig:perf_spec_llama} and Figure~\ref{fig:perf_spec_deepseek} show that Falcon consistently outperforms the other serial generation approaches when evaluated in 6 different datasets. The gain from generating draft tree in parallel comes from the observation that both the draft model and target model could not have significant inference speed improvement from using 8 GPUs instead of 4 GPUs. Therefore, if we allocate 8 GPUs instead of 4 GPUs for both models, the effect is close to just use 4 GPUs while letting the other 4 idle. On the other hand, our approach, allocating 4 GPUs for each of the models, could better utilize the GPUs and provide higher token generation speed.

% The tradeoff that our approach has to face is that, when the verified token goes out of the generated draft tree, the target model will have to wait for the draft model to generate a candidate tree large enough to do. We run experiment shows that..... (the amount of time misprediction occured) and the compression ratio of our approach is only marginally smaller under the same batch size, making the whole approach an improvement to the original sequential sepeculative decoding methods. 

\subsection{Latency-optimized operator microbenchmark}
\label{sec:eval_ll_kernels}
% Please add the following required packages to your document preamble:
% \usepackage{graphicx}
% To show how 

\begin{table*}[]
\resizebox{\textwidth}{!}{%
\begin{tabular}{c|ccc|ccc|ccc|ccc|ccc}
 &
  \multicolumn{3}{c|}{Fused GEMM (attn)}  &  \multicolumn{3}{c|}{Fused GEMM (mlp)}&
  \multicolumn{3}{c|}{SwiGLU} &
  \multicolumn{3}{c|}{Attn, len(context)=500}&
  \multicolumn{3}{c}{Attn, len(context)=1000}\\ \hline
            & \makebox[0.5cm]{Ours}& \makebox[0.5cm]{vllm}& \makebox[0.5cm]{TRT}& \makebox[0.5cm]{Ours}& \makebox[0.5cm]{vllm}&\makebox[0.5cm]{TRT}& 
              \makebox[0.5cm]{Ours}& \makebox[0.5cm]{vllm}& \makebox[0.5cm]{TRT}& 
              \makebox[0.5cm]{Ours}& \makebox[0.5cm]{FI}&\makebox[0.5cm]{FA}& 
              \makebox[0.5cm]{Ours}& \makebox[0.5cm]{FI}&\makebox[0.5cm]{FA}\\ \hline
1B, tp = 4  &   \textbf{5.9us}&   12.9us&   10.2us& \textbf{8.3us}& 18.8us&11.0us&   \textbf{5.8us}&   12.3us&   11.5us &   \textbf{6.3us}&    13.6us&13.8us&   \textbf{6.2us}&    22.8us&13.8us\\
3B, tp = 4  &   \textbf{6.4us}&   12.3us&   10.0us& \textbf{8.7us}& 17.0us&10.3us&   \textbf{7.2us}&   12.5us&   11.8us &   \textbf{6.2us}&    19.1us&17.2us&   \textbf{8.0us}&    33.4us&18.3us\\
8B, tp = 4  &   \textbf{7.7us}&   14.6us&   10.3us& \textbf{10.5us}& 16.5us&11.9us&   \textbf{15.0us}&   16.3us&   15us &   \textbf{6.4us}&    19.4us&17.75us&   \textbf{8.1us}&    33.4us&18.13us\\
70B, tp = 4 &   \textbf{11.72us}&   16.9us&   15.5us& \textbf{24.3us}& 25.4us&26.1us&   49us&   36.7us&   \textbf{31.9us} &   \textbf{9.6us}&    19.2us&19.0us&   \textbf{13.4us}&    33.2us&19.1us\\
70B, tp = 8 &   \textbf{13.3us}&   25.7us&   17.2us& \textbf{19.9us}& 29.5us&20.7us&   23.6us&   22.5us&   \textbf{22us} &   \textbf{6.4us}&    18.6us&17.8us&   \textbf{8.1us}&    32.7us&18.2us\end{tabular}%
}
\caption{Time per kernel of our optimized operator under batch size 8 across different TP configurations in the Llama family.}
\vspace{-0.1in}
\label{tab:kernel_runtime}
\end{table*}

% \zy{compare the kernel time of gemm all reduce, attention, swiglu}

In Table \ref{tab:kernel_runtime}, we show the individual kernel times of our latency-optimized kernels and kernels in other baselines. We focus on the Llama3 model family as a specific example, while our optimizations are also applied to other model families in our benchmark, achieving similar speedup.

\textbf{Fused GEMM with all reduce} In each layer of the Llama3 model under tensor parallelism, there are two all-reduce operations, one in the attention block, and the other in the MLP block. We fuse each all-reduce operation with the previous GEMM. In the implementation of both vllm and TRTLLM, they use their implementation of one-shot all-reduce, and they perform the all-reduce operations after the GEMM operations as separate kernels. We sum up the time spent on two kernels as their total time. In our solution, we fuse the GEMM operation with our implementation of one-shot all-reduce. As shown in Table~\ref{tab:kernel_runtime}, our fused GEMM-all-reduce kernel consistently outperforms vllm and TRTLLM on all five model configurations. Our improvement is larger when the amount of compute is less: this happens for the fused GEMM in attention block for all models (with a time reduction of $23\%$-$43\%$) and that in MLP block for the smaller models, 1B, 3B, and 8B (with a time reduction of $16\%$-$25\%$).

\begin{figure}[t]
    \centering
    \includegraphics[width=\linewidth]{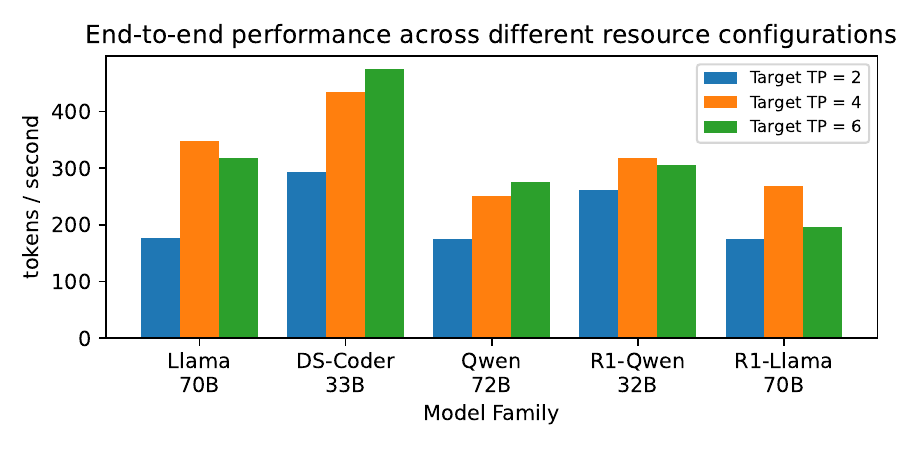}
    \vspace{-0.4in}
    \caption{End-to-end decoding speed comparison using different resource configurations for draft model and target model}
    \vspace{-0.20in}
    \label{fig:perf_gpu_allocation}
\end{figure}
\textbf{Attention operator} For the attention operator, we compare our implementation with two popular attention libraries: FlashAttention (FA) and FlashInfer (FI). To make a fair comparison to FA and FI, which only support square kernels, during the benchmark, we only feed square masks to our tree attention kernel, even though it supports a more general non-square mask. We use BatchPrefillWithPagedKVCacheWrapper in FlashInfer, which supports a square custom mask. However, the kernel is not optimized for small numbers of attention heads, and, therefore, performs much worse than the other baseline, FA. FA uses kernel \textit{flash\_fwd\_splitkv\_kernel} to split each kv head across different threadblocks to compute the attention score and sum faster, and then it launches another kernel \textit{flash\_fwd\_splitkv\_combine\_kernel} to aggregate the results across different threadblocks. On the other hand, our kernel fuses those two kernels into one, using the NCCL LL protocol to perform an implicit barrier, saving time for the overhead of synchronization and an extra kernel launch. As shown in Table~\ref{tab:kernel_runtime}, our optimization kernel constantly saves $30\%$ to $56\%$ of the running time compared to FA under two representative context lengths, 500 and 1000, across different model configurations.

\textbf{SwiGLU operator} In each layer of the Llama3 model, we fuse the four operations in the SwiGLU operator ($SwiGLU(x, $ $W, V, b, c) = \sigma (xW + b) \oplus (xV + c)$) into one, reducing data movement and extra kernel launches. Both vllm and TRTLLM fuse the dot product operator with the $\sigma$ activation operator. Furthermore, vllm fuses the first two matrix multiplications into one operator. As shown in Table~\ref{tab:kernel_runtime}, our SwiGLU optimization outperforms other baselines (with a time reduction of $39\%$-$50\%$) when the model is small (1B, 3B), where latency dominates the whole execution time. When the model is larger and the amount of computation is more (for the 70B model), TRTLLM and vllm outperform our kernel since they have more optimized kernels (including AWQ marlin, where a more intricate layout of weight matrices is used to improve performance), while our kernel is based on simple tile-based GEMM and uniform layout. 

% \zy{For attention operator, maybe just show 3 input lenght? (500, 1000, 2000) and put them all into one table with other kernels to save space }

\subsection{Design choices}
% this should cover more specfic things like (CPU overhead, compression ratio, etc)
\label{sec:eval_design_choices}
\textbf{Choice about target batch size, draft batch size, depth} In section~\ref{sec:motivation_spec}, Figure~\ref{fig:motivation_batchsize} shows that increasing the draft model batch size over 8 only increases the compression ratio marginally. For the target model, there is still an average increase of $24.5\%$  when we increase the batch size from 8 to 16. However, for example, if we increase the batch size of the Llama3-70B model from 8 to 16 under TP=4, the inference time will increase from $10.39$ ms to $13.25$ ms (by $29\%$). Thus, the increase in compression ratio does not cover the increase in inference time. Furthermore, reducing the batch size to under 8 does not reduce the inference time (e.g. 
It takes $10.32$ms under batch size 4) since the smallest first input dimension of the matrix multiplication tensor core operation is 8. As a result, we use batch size = 8 for both our target model and draft model.

To choose $d$, the number of tree expansions each round, as reasoned in $\S\ref{sec:parellel_tree_generation}$, we choose one of the two integers that are closest to the time ratio of a target model inference and one round of draft tree expansion. In the benchmark, for each model pair, we run \paper{} using those two different $d$s based on the draft and target model execution time and choose the configuration with the higher decoding speed.

% \zy{summarize why we use target batch size = 8, draft batch size = 8, depth = target model time / draft model time. Support using exp results}

% \textbf{Compression ratio, total running time} across different batch size (e.g. 8, 16, 32), depth = 4,5,6 (for eagle/serial).  
\textbf{Draft model and target model resource allocation.}  Figure~\ref{fig:perf_gpu_allocation}  illustrates how flexible allocation of GPU resources between draft and target models affects overall decoding performance. Specifically, it shows the performance of each model family when we allocate different numbers of GPUs to the target and draft models. We only consider TP=2,4,6 for the target model since even degree of tensor parallelism is more aligned to the attention operators and matrix multiplications and thus requires less padding.  For Deepseek-Coder 33B and Qwen2-72B, we find it optimal to use TP=6 for the target model and TP=2 for the draft model, and for other models, it is the best to use TP=4 for both models since for those model pairs, giving more compute to a more capable draft model can increase the number of tree layers and thus increase the number of accepted tokens per round. 
% \zy{maybe should back this up using compression ratio data}

% \textbf{Greedy decoding, sampling?}

% \textbf{CPU overhead} We benchmark the overhead of our draft tree manage

% \textbf{Prefill speed?} Show that our inference does not change prefill speed too much, or it only occupies a small portion of the time.

% \textbf{Effect of changing batch size $bs$, depth of draft tree $d$, width of draft tree $w$, number of children $c$ to consider for each node in draft tree}

% \textbf{Time breakdown for each stage (portion of time that we need to wait for new layers.)}

% \textbf{Fall out ratio, }

% \textbf{Comparison with traditional speculative decoding}  \zy{compare with eagle and our serial version}

\section{Discussion and Limitations}

\textbf{EAGLE speculative decoding vs smaller-draft speculative decoding} While \paper{} could make normal speculative decoding with a smaller draft model faster, it is non-trivial to apply them into EAGLE speculative decoding, since the EAGLE model takes the target model output of the verified tokens as input, imposing a even stricter dependency between draft model and target model that is hard to break. Therefore, our approach is more suitable when an "independent" draft model, preferably a smaller model from the same model family, is available. How to integrate our method with EAGLE is left as future work.

\noindent \textbf{High-throughput serving}
Our use of int4 quantized model and its latency-optimized kernels are naturally not suitable for high-throughput serving scenarios where the batch size is larger (e.g. $\geq 64$). However, we believe our parallel tree generation algorithm could still be useful for larger batch sizes (especially when the GPU interconnect is slow) since it effectively decreases the amount of communication across different GPUs. However, as speculative decoding is generally useful when the workload of the system is relatively light \cite{liu2024optimizingspeculativedecodingserving}, when the system workload is high and it is more cost-efficient to run autogressive decoding, our approach, based on speculative decoding, becomes less effective as well.
% \zy{add observation that under int4 TP, the inference time of draft model become closer to that of target model, and thus the batch size of both model should be smaller since the marginal cost of increasing batch decreases a lot }

\noindent \textbf{Disaggregated execution} LLM serving systems such as \cite{distserve, patel2024splitwiseefficientgenerativellm} use the idea of disaggregating the prefill phase and the decoding phase. \paper's parallel tree generation idea is similar in spirit to those approaches, allowing different types of computation to scale independently. On the other hand, \paper{} focuses on disaggregating different components in the decoding phase and is thus orthogonal to those approaches. Therefore, approaches that accelerate the prefill phase \cite{liu2024cachegenkvcachecompression, zhao2024prepackingsimplemethodfast} are complementary to our system.

% \noindent \textbf{Running draft model and the target model in parallel}
% There is concurrent work like \cite{liu2024parallelspeculativedecodingadaptive}, which runs the draft model and target model in parallel. However, such an approach only considers sequence-based speculative coding, which is slower than tree-based methods because of the lower successful guess ratio. On the other hand, we design a KV cache management scheme and enable the system to run tree-based speculative decoding in parallel, yielding faster decoding speed.

% PipeInfer \cite{butler2024pipeinferacceleratingllminference} runs target model using pipeline parallelism, which leaves a large amount of bubbles (CPU idling time) in the systems, wasting compute resource. Besides, PipeInfer keeps multiple runs of speculation, which could potentially calculate the same speculation at the same time, while our work keeps only one copy of the tree and thus not generating duplicate speculations.

\section{Conclusion}

Increasing the decoding speed of large language models by scaling speculative decoding is important yet challenging for reducing end-to-end request latency. This paper proposes \paper{}, which includes three key innovations: \textit{parallel tree generation}, \textit{consistent KV cache management}, and \textit{latency-optimized kernels}. Combining those three techniques, we are able to serve five different models in different model families with an average decoding speed $1.75 \times$ faster than the most competitive baselines. Particularly, \paper{} can serve the Llama3 70B model on an average of 348 tokens per second, which is currently the fastest system using Nvidia Hopper GPUs. We believe the parallel tree generation with KV cache consistency and fused operator strategies are general and independent of the particular model families we test, so we expect these key techniques \paper{} to generalize to future speculative decoding architectures.

\clearpage

\bibliographystyle{plainnat}
\bibliography{main}

\begin{thebibliography}{46}
\providecommand{\natexlab}[1]{#1}
\providecommand{\url}[1]{\texttt{#1}}
\expandafter\ifx\csname urlstyle\endcsname\relax
  \providecommand{\doi}[1]{doi: #1}\else
  \providecommand{\doi}{doi: \begingroup \urlstyle{rm}\Url}\fi

\bibitem[Agrawal et~al.(2024)Agrawal, Kedia, Panwar, Mohan, Kwatra, Gulavani, Tumanov, and Ramjee]{sarathichunkedprefill}
Amey Agrawal, Nitin Kedia, Ashish Panwar, Jayashree Mohan, Nipun Kwatra, Bhargav Gulavani, Alexey Tumanov, and Ramachandran Ramjee.
\newblock Taming {Throughput-Latency} tradeoff in {LLM} inference with {Sarathi-Serve}.
\newblock In \emph{18th USENIX Symposium on Operating Systems Design and Implementation (OSDI 24)}, pages 117--134, Santa Clara, CA, July 2024. USENIX Association.
\newblock ISBN 978-1-939133-40-3.
\newblock URL \url{https://www.usenix.org/conference/osdi24/presentation/agrawal}.

\bibitem[Butler et~al.(2024)Butler, Yu, Mazaheri, and Jannesari]{butler2024pipeinferacceleratingllminference}
Branden Butler, Sixing Yu, Arya Mazaheri, and Ali Jannesari.
\newblock Pipeinfer: Accelerating llm inference using asynchronous pipelined speculation, 2024.
\newblock URL \url{https://arxiv.org/abs/2407.11798}.

\bibitem[Cai et~al.(2024)Cai, Li, Geng, Peng, Lee, Chen, and Dao]{cai2024medusasimplellminference}
Tianle Cai, Yuhong Li, Zhengyang Geng, Hongwu Peng, Jason~D. Lee, Deming Chen, and Tri Dao.
\newblock Medusa: Simple llm inference acceleration framework with multiple decoding heads, 2024.
\newblock URL \url{https://arxiv.org/abs/2401.10774}.

\bibitem[Caramancion(2024)]{search2_caramancion2024largelanguagemodelsvs}
Kevin~Matthe Caramancion.
\newblock Large language models vs. search engines: Evaluating user preferences across varied information retrieval scenarios, 2024.
\newblock URL \url{https://arxiv.org/abs/2401.05761}.

\bibitem[Chen et~al.(2023)Chen, Borgeaud, Irving, Lespiau, Sifre, and Jumper]{chen2023acceleratinglargelanguagemodel}
Charlie Chen, Sebastian Borgeaud, Geoffrey Irving, Jean-Baptiste Lespiau, Laurent Sifre, and John Jumper.
\newblock Accelerating large language model decoding with speculative sampling, 2023.
\newblock URL \url{https://arxiv.org/abs/2302.01318}.

\bibitem[Chen et~al.(2021{\natexlab{a}})Chen, Tworek, Jun, Yuan, de~Oliveira~Pinto, Kaplan, Edwards, Burda, Joseph, Brockman, Ray, Puri, Krueger, Petrov, Khlaaf, Sastry, Mishkin, Chan, Gray, Ryder, Pavlov, Power, Kaiser, Bavarian, Winter, Tillet, Such, Cummings, Plappert, Chantzis, Barnes, Herbert-Voss, Guss, Nichol, Paino, Tezak, Tang, Babuschkin, Balaji, Jain, Saunders, Hesse, Carr, Leike, Achiam, Misra, Morikawa, Radford, Knight, Brundage, Murati, Mayer, Welinder, McGrew, Amodei, McCandlish, Sutskever, and Zaremba]{coder2chen2021evaluatinglargelanguagemodels}
Mark Chen, Jerry Tworek, Heewoo Jun, Qiming Yuan, Henrique~Ponde de~Oliveira~Pinto, Jared Kaplan, Harri Edwards, Yuri Burda, Nicholas Joseph, Greg Brockman, Alex Ray, Raul Puri, Gretchen Krueger, Michael Petrov, Heidy Khlaaf, Girish Sastry, Pamela Mishkin, Brooke Chan, Scott Gray, Nick Ryder, Mikhail Pavlov, Alethea Power, Lukasz Kaiser, Mohammad Bavarian, Clemens Winter, Philippe Tillet, Felipe~Petroski Such, Dave Cummings, Matthias Plappert, Fotios Chantzis, Elizabeth Barnes, Ariel Herbert-Voss, William~Hebgen Guss, Alex Nichol, Alex Paino, Nikolas Tezak, Jie Tang, Igor Babuschkin, Suchir Balaji, Shantanu Jain, William Saunders, Christopher Hesse, Andrew~N. Carr, Jan Leike, Josh Achiam, Vedant Misra, Evan Morikawa, Alec Radford, Matthew Knight, Miles Brundage, Mira Murati, Katie Mayer, Peter Welinder, Bob McGrew, Dario Amodei, Sam McCandlish, Ilya Sutskever, and Wojciech Zaremba.
\newblock Evaluating large language models trained on code, 2021{\natexlab{a}}.
\newblock URL \url{https://arxiv.org/abs/2107.03374}.

\bibitem[Chen et~al.(2021{\natexlab{b}})Chen, Tworek, Jun, Yuan, de~Oliveira~Pinto, Kaplan, Edwards, Burda, Joseph, Brockman, Ray, Puri, Krueger, Petrov, Khlaaf, Sastry, Mishkin, Chan, Gray, Ryder, Pavlov, Power, Kaiser, Bavarian, Winter, Tillet, Such, Cummings, Plappert, Chantzis, Barnes, Herbert-Voss, Guss, Nichol, Paino, Tezak, Tang, Babuschkin, Balaji, Jain, Saunders, Hesse, Carr, Leike, Achiam, Misra, Morikawa, Radford, Knight, Brundage, Murati, Mayer, Welinder, McGrew, Amodei, McCandlish, Sutskever, and Zaremba]{chen2021evaluatinglargelanguagemodelshumaneval}
Mark Chen, Jerry Tworek, Heewoo Jun, Qiming Yuan, Henrique~Ponde de~Oliveira~Pinto, Jared Kaplan, Harri Edwards, Yuri Burda, Nicholas Joseph, Greg Brockman, Alex Ray, Raul Puri, Gretchen Krueger, Michael Petrov, Heidy Khlaaf, Girish Sastry, Pamela Mishkin, Brooke Chan, Scott Gray, Nick Ryder, Mikhail Pavlov, Alethea Power, Lukasz Kaiser, Mohammad Bavarian, Clemens Winter, Philippe Tillet, Felipe~Petroski Such, Dave Cummings, Matthias Plappert, Fotios Chantzis, Elizabeth Barnes, Ariel Herbert-Voss, William~Hebgen Guss, Alex Nichol, Alex Paino, Nikolas Tezak, Jie Tang, Igor Babuschkin, Suchir Balaji, Shantanu Jain, William Saunders, Christopher Hesse, Andrew~N. Carr, Jan Leike, Josh Achiam, Vedant Misra, Evan Morikawa, Alec Radford, Matthew Knight, Miles Brundage, Mira Murati, Katie Mayer, Peter Welinder, Bob McGrew, Dario Amodei, Sam McCandlish, Ilya Sutskever, and Wojciech Zaremba.
\newblock Evaluating large language models trained on code, 2021{\natexlab{b}}.
\newblock URL \url{https://arxiv.org/abs/2107.03374}.

\bibitem[Cobbe et~al.(2021)Cobbe, Kosaraju, Bavarian, Chen, Jun, Kaiser, Plappert, Tworek, Hilton, Nakano, Hesse, and Schulman]{cobbe2021trainingverifierssolvemathgsm8k}
Karl Cobbe, Vineet Kosaraju, Mohammad Bavarian, Mark Chen, Heewoo Jun, Lukasz Kaiser, Matthias Plappert, Jerry Tworek, Jacob Hilton, Reiichiro Nakano, Christopher Hesse, and John Schulman.
\newblock Training verifiers to solve math word problems, 2021.
\newblock URL \url{https://arxiv.org/abs/2110.14168}.

\bibitem[DeepSeek-AI et~al.(2025)DeepSeek-AI, Guo, Yang, Zhang, Song, Zhang, Xu, Zhu, Ma, Wang, Bi, Zhang, Yu, Wu, Wu, Gou, Shao, Li, Gao, Liu, Xue, Wang, Wu, Feng, Lu, Zhao, Deng, Zhang, Ruan, Dai, Chen, Ji, Li, Lin, Dai, Luo, Hao, Chen, Li, Zhang, Bao, Xu, Wang, Ding, Xin, Gao, Qu, Li, Guo, Li, Wang, Chen, Yuan, Qiu, Li, Cai, Ni, Liang, Chen, Dong, Hu, Gao, Guan, Huang, Yu, Wang, Zhang, Zhao, Wang, Zhang, Xu, Xia, Zhang, Zhang, Tang, Li, Wang, Li, Tian, Huang, Zhang, Wang, Chen, Du, Ge, Zhang, Pan, Wang, Chen, Jin, Chen, Lu, Zhou, Chen, Ye, Wang, Yu, Zhou, Pan, Li, Zhou, Wu, Ye, Yun, Pei, Sun, Wang, Zeng, Zhao, Liu, Liang, Gao, Yu, Zhang, Xiao, An, Liu, Wang, Chen, Nie, Cheng, Liu, Xie, Liu, Yang, Li, Su, Lin, Li, Jin, Shen, Chen, Sun, Wang, Song, Zhou, Wang, Shan, Li, Wang, Wei, Zhang, Xu, Li, Zhao, Sun, Wang, Yu, Zhang, Shi, Xiong, He, Piao, Wang, Tan, Ma, Liu, Guo, Ou, Wang, Gong, Zou, He, Xiong, Luo, You, Liu, Zhou, Zhu, Xu, Huang, Li, Zheng, Zhu, Ma, Tang, Zha, Yan, Ren, Ren, Sha, Fu, Xu, Xie, Zhang,
  Hao, Ma, Yan, Wu, Gu, Zhu, Liu, Li, Xie, Song, Pan, Huang, Xu, Zhang, and Zhang]{deepseekai2025deepseekr1incentivizingreasoningcapability}
DeepSeek-AI, Daya Guo, Dejian Yang, Haowei Zhang, Junxiao Song, Ruoyu Zhang, Runxin Xu, Qihao Zhu, Shirong Ma, Peiyi Wang, Xiao Bi, Xiaokang Zhang, Xingkai Yu, Yu~Wu, Z.~F. Wu, Zhibin Gou, Zhihong Shao, Zhuoshu Li, Ziyi Gao, Aixin Liu, Bing Xue, Bingxuan Wang, Bochao Wu, Bei Feng, Chengda Lu, Chenggang Zhao, Chengqi Deng, Chenyu Zhang, Chong Ruan, Damai Dai, Deli Chen, Dongjie Ji, Erhang Li, Fangyun Lin, Fucong Dai, Fuli Luo, Guangbo Hao, Guanting Chen, Guowei Li, H.~Zhang, Han Bao, Hanwei Xu, Haocheng Wang, Honghui Ding, Huajian Xin, Huazuo Gao, Hui Qu, Hui Li, Jianzhong Guo, Jiashi Li, Jiawei Wang, Jingchang Chen, Jingyang Yuan, Junjie Qiu, Junlong Li, J.~L. Cai, Jiaqi Ni, Jian Liang, Jin Chen, Kai Dong, Kai Hu, Kaige Gao, Kang Guan, Kexin Huang, Kuai Yu, Lean Wang, Lecong Zhang, Liang Zhao, Litong Wang, Liyue Zhang, Lei Xu, Leyi Xia, Mingchuan Zhang, Minghua Zhang, Minghui Tang, Meng Li, Miaojun Wang, Mingming Li, Ning Tian, Panpan Huang, Peng Zhang, Qiancheng Wang, Qinyu Chen, Qiushi Du, Ruiqi Ge, Ruisong
  Zhang, Ruizhe Pan, Runji Wang, R.~J. Chen, R.~L. Jin, Ruyi Chen, Shanghao Lu, Shangyan Zhou, Shanhuang Chen, Shengfeng Ye, Shiyu Wang, Shuiping Yu, Shunfeng Zhou, Shuting Pan, S.~S. Li, Shuang Zhou, Shaoqing Wu, Shengfeng Ye, Tao Yun, Tian Pei, Tianyu Sun, T.~Wang, Wangding Zeng, Wanjia Zhao, Wen Liu, Wenfeng Liang, Wenjun Gao, Wenqin Yu, Wentao Zhang, W.~L. Xiao, Wei An, Xiaodong Liu, Xiaohan Wang, Xiaokang Chen, Xiaotao Nie, Xin Cheng, Xin Liu, Xin Xie, Xingchao Liu, Xinyu Yang, Xinyuan Li, Xuecheng Su, Xuheng Lin, X.~Q. Li, Xiangyue Jin, Xiaojin Shen, Xiaosha Chen, Xiaowen Sun, Xiaoxiang Wang, Xinnan Song, Xinyi Zhou, Xianzu Wang, Xinxia Shan, Y.~K. Li, Y.~Q. Wang, Y.~X. Wei, Yang Zhang, Yanhong Xu, Yao Li, Yao Zhao, Yaofeng Sun, Yaohui Wang, Yi~Yu, Yichao Zhang, Yifan Shi, Yiliang Xiong, Ying He, Yishi Piao, Yisong Wang, Yixuan Tan, Yiyang Ma, Yiyuan Liu, Yongqiang Guo, Yuan Ou, Yuduan Wang, Yue Gong, Yuheng Zou, Yujia He, Yunfan Xiong, Yuxiang Luo, Yuxiang You, Yuxuan Liu, Yuyang Zhou, Y.~X. Zhu,
  Yanhong Xu, Yanping Huang, Yaohui Li, Yi~Zheng, Yuchen Zhu, Yunxian Ma, Ying Tang, Yukun Zha, Yuting Yan, Z.~Z. Ren, Zehui Ren, Zhangli Sha, Zhe Fu, Zhean Xu, Zhenda Xie, Zhengyan Zhang, Zhewen Hao, Zhicheng Ma, Zhigang Yan, Zhiyu Wu, Zihui Gu, Zijia Zhu, Zijun Liu, Zilin Li, Ziwei Xie, Ziyang Song, Zizheng Pan, Zhen Huang, Zhipeng Xu, Zhongyu Zhang, and Zhen Zhang.
\newblock Deepseek-r1: Incentivizing reasoning capability in llms via reinforcement learning, 2025.
\newblock URL \url{https://arxiv.org/abs/2501.12948}.

\bibitem[Fan et~al.(2020)Fan, Rong, Meng, Cao, Wang, Zheng, Wu, Long, Yang, Xia, Diao, Liu, and Lin]{fan2020dapplepipelineddataparallel}
Shiqing Fan, Yi~Rong, Chen Meng, Zongyan Cao, Siyu Wang, Zhen Zheng, Chuan Wu, Guoping Long, Jun Yang, Lixue Xia, Lansong Diao, Xiaoyong Liu, and Wei Lin.
\newblock Dapple: A pipelined data parallel approach for training large models, 2020.
\newblock URL \url{https://arxiv.org/abs/2007.01045}.

\bibitem[Guo et~al.(2023)Guo, Xu, Duan, Yin, and McAuley]{coder1guo2023longcoderlongrangepretrainedlanguage}
Daya Guo, Canwen Xu, Nan Duan, Jian Yin, and Julian McAuley.
\newblock Longcoder: A long-range pre-trained language model for code completion, 2023.
\newblock URL \url{https://arxiv.org/abs/2306.14893}.

\bibitem[Guo et~al.(2024)Guo, Zhu, Yang, Xie, Dong, Zhang, Chen, Bi, Wu, Li, Luo, Xiong, and Liang]{guo2024deepseekcoderlargelanguagemodel}
Daya Guo, Qihao Zhu, Dejian Yang, Zhenda Xie, Kai Dong, Wentao Zhang, Guanting Chen, Xiao Bi, Y.~Wu, Y.~K. Li, Fuli Luo, Yingfei Xiong, and Wenfeng Liang.
\newblock Deepseek-coder: When the large language model meets programming -- the rise of code intelligence, 2024.
\newblock URL \url{https://arxiv.org/abs/2401.14196}.

\bibitem[Hariri(2025)]{hariri2025unlockingpotentialchatgptcomprehensive}
Walid Hariri.
\newblock Unlocking the potential of chatgpt: A comprehensive exploration of its applications, advantages, limitations, and future directions in natural language processing, 2025.
\newblock URL \url{https://arxiv.org/abs/2304.02017}.

\bibitem[Huang et~al.(2019)Huang, Cheng, Bapna, Firat, Chen, Chen, Lee, Ngiam, Le, Wu, and Chen]{huang2019gpipeefficienttraininggiant}
Yanping Huang, Youlong Cheng, Ankur Bapna, Orhan Firat, Mia~Xu Chen, Dehao Chen, HyoukJoong Lee, Jiquan Ngiam, Quoc~V. Le, Yonghui Wu, and Zhifeng Chen.
\newblock Gpipe: Efficient training of giant neural networks using pipeline parallelism, 2019.
\newblock URL \url{https://arxiv.org/abs/1811.06965}.

\bibitem[Izadi et~al.(2024)Izadi, Katzy, van Dam, Otten, Popescu, and van Deursen]{izadi2024languagemodelscodecompletion}
Maliheh Izadi, Jonathan Katzy, Tim van Dam, Marc Otten, Razvan~Mihai Popescu, and Arie van Deursen.
\newblock Language models for code completion: A practical evaluation, 2024.
\newblock URL \url{https://arxiv.org/abs/2402.16197}.

\bibitem[Kwiatkowski et~al.(2019)Kwiatkowski, Palomaki, Redfield, Collins, Parikh, Alberti, Epstein, Polosukhin, Devlin, Lee, Toutanova, Jones, Kelcey, Chang, Dai, Uszkoreit, Le, and Petrov]{kwiatkowski-etal-2019-naturalquestions}
Tom Kwiatkowski, Jennimaria Palomaki, Olivia Redfield, Michael Collins, Ankur Parikh, Chris Alberti, Danielle Epstein, Illia Polosukhin, Jacob Devlin, Kenton Lee, Kristina Toutanova, Llion Jones, Matthew Kelcey, Ming-Wei Chang, Andrew~M. Dai, Jakob Uszkoreit, Quoc Le, and Slav Petrov.
\newblock Natural questions: A benchmark for question answering research.
\newblock \emph{Transactions of the Association for Computational Linguistics}, 7:\penalty0 452--466, 2019.
\newblock \doi{10.1162/tacl_a_00276}.
\newblock URL \url{https://aclanthology.org/Q19-1026/}.

\bibitem[Kwon et~al.(2023)Kwon, Li, Zhuang, Sheng, Zheng, Yu, Gonzalez, Zhang, and Stoica]{vllmkwon2023efficientmemorymanagementlarge}
Woosuk Kwon, Zhuohan Li, Siyuan Zhuang, Ying Sheng, Lianmin Zheng, Cody~Hao Yu, Joseph~E. Gonzalez, Hao Zhang, and Ion Stoica.
\newblock Efficient memory management for large language model serving with pagedattention, 2023.
\newblock URL \url{https://arxiv.org/abs/2309.06180}.

\bibitem[Leviathan et~al.(2023)Leviathan, Kalman, and Matias]{leviathan2023fastinferencetransformersspeculative}
Yaniv Leviathan, Matan Kalman, and Yossi Matias.
\newblock Fast inference from transformers via speculative decoding, 2023.
\newblock URL \url{https://arxiv.org/abs/2211.17192}.

\bibitem[Li et~al.(2024)Li, Wei, Zhang, and Zhang]{li2024eagle2fasterinferencelanguage}
Yuhui Li, Fangyun Wei, Chao Zhang, and Hongyang Zhang.
\newblock Eagle-2: Faster inference of language models with dynamic draft trees, 2024.
\newblock URL \url{https://arxiv.org/abs/2406.16858}.

\bibitem[Lin et~al.(2024)Lin, Tang, Tang, Yang, Chen, Wang, Xiao, Dang, Gan, and Han]{lin2024awqactivationawareweightquantization}
Ji~Lin, Jiaming Tang, Haotian Tang, Shang Yang, Wei-Ming Chen, Wei-Chen Wang, Guangxuan Xiao, Xingyu Dang, Chuang Gan, and Song Han.
\newblock Awq: Activation-aware weight quantization for llm compression and acceleration, 2024.
\newblock URL \url{https://arxiv.org/abs/2306.00978}.

\bibitem[Liu et~al.(2024{\natexlab{a}})Liu, Li, Lv, Liu, Zhu, and Hu]{liu2024parallelspeculativedecodingadaptive}
Tianyu Liu, Yun Li, Qitan Lv, Kai Liu, Jianchen Zhu, and Winston Hu.
\newblock Parallel speculative decoding with adaptive draft length, 2024{\natexlab{a}}.
\newblock URL \url{https://arxiv.org/abs/2408.11850}.

\bibitem[Liu et~al.(2024{\natexlab{b}})Liu, Daniel, Hu, Kwon, Li, Mo, Cheung, Deng, Stoica, and Zhang]{liu2024optimizingspeculativedecodingserving}
Xiaoxuan Liu, Cade Daniel, Langxiang Hu, Woosuk Kwon, Zhuohan Li, Xiangxi Mo, Alvin Cheung, Zhijie Deng, Ion Stoica, and Hao Zhang.
\newblock Optimizing speculative decoding for serving large language models using goodput, 2024{\natexlab{b}}.
\newblock URL \url{https://arxiv.org/abs/2406.14066}.

\bibitem[Liu et~al.(2024{\natexlab{c}})Liu, Li, Cheng, Ray, Huang, Zhang, Du, Yao, Lu, Ananthanarayanan, Maire, Hoffmann, Holtzman, and Jiang]{liu2024cachegenkvcachecompression}
Yuhan Liu, Hanchen Li, Yihua Cheng, Siddhant Ray, Yuyang Huang, Qizheng Zhang, Kuntai Du, Jiayi Yao, Shan Lu, Ganesh Ananthanarayanan, Michael Maire, Henry Hoffmann, Ari Holtzman, and Junchen Jiang.
\newblock Cachegen: Kv cache compression and streaming for fast large language model serving, 2024{\natexlab{c}}.
\newblock URL \url{https://arxiv.org/abs/2310.07240}.

\bibitem[McDanel(2024)]{mcdanel2024amusdasynchronousmultidevicespeculative}
Bradley McDanel.
\newblock Amusd: Asynchronous multi-device speculative decoding for llm acceleration, 2024.
\newblock URL \url{https://arxiv.org/abs/2410.17375}.

\bibitem[Miao et~al.(2024)Miao, Oliaro, Zhang, Cheng, Wang, Zhang, Wong, Zhu, Yang, Shi, Shi, Chen, Arfeen, Abhyankar, and Jia]{Miao_2024}
Xupeng Miao, Gabriele Oliaro, Zhihao Zhang, Xinhao Cheng, Zeyu Wang, Zhengxin Zhang, Rae Ying~Yee Wong, Alan Zhu, Lijie Yang, Xiaoxiang Shi, Chunan Shi, Zhuoming Chen, Daiyaan Arfeen, Reyna Abhyankar, and Zhihao Jia.
\newblock Specinfer: Accelerating large language model serving with tree-based speculative inference and verification.
\newblock In \emph{Proceedings of the 29th ACM International Conference on Architectural Support for Programming Languages and Operating Systems, Volume 3}, ASPLOS ’24. ACM, April 2024.
\newblock \doi{10.1145/3620666.3651335}.
\newblock URL \url{http://dx.doi.org/10.1145/3620666.3651335}.

\bibitem[Nallapati et~al.(2016)Nallapati, Zhou, dos santos, Gulcehre, and Xiang]{nallapati2016abstractivetextsummarizationusingcnndaily}
Ramesh Nallapati, Bowen Zhou, Cicero~Nogueira dos santos, Caglar Gulcehre, and Bing Xiang.
\newblock Abstractive text summarization using sequence-to-sequence rnns and beyond, 2016.
\newblock URL \url{https://arxiv.org/abs/1602.06023}.

\bibitem[Nvidia()]{trtllm}
Nvidia.
\newblock Nvidia/tensorrt-llm: A tensorrt toolbox for optimized large language model inference.
\newblock \url{https://github.com/NVIDIA/TensorRT-LLM}.

\bibitem[OpenAI(2024)]{openai2024gpt4technicalreport}
OpenAI.
\newblock Gpt-4 technical report, 2024.
\newblock URL \url{https://arxiv.org/abs/2303.08774}.

\bibitem[Patel et~al.(2024)Patel, Choukse, Zhang, Shah, Íñigo Goiri, Maleki, and Bianchini]{patel2024splitwiseefficientgenerativellm}
Pratyush Patel, Esha Choukse, Chaojie Zhang, Aashaka Shah, Íñigo Goiri, Saeed Maleki, and Ricardo Bianchini.
\newblock Splitwise: Efficient generative llm inference using phase splitting, 2024.
\newblock URL \url{https://arxiv.org/abs/2311.18677}.

\bibitem[Shoeybi et~al.(2020)Shoeybi, Patwary, Puri, LeGresley, Casper, and Catanzaro]{shoeybi2020megatronlmtrainingmultibillionparameter}
Mohammad Shoeybi, Mostofa Patwary, Raul Puri, Patrick LeGresley, Jared Casper, and Bryan Catanzaro.
\newblock Megatron-lm: Training multi-billion parameter language models using model parallelism, 2020.
\newblock URL \url{https://arxiv.org/abs/1909.08053}.

\bibitem[Tatsu-Lab()]{alpaca}
Tatsu-Lab.
\newblock Tatsu-lab/stanford-alpaca: Code and documentation to train stanford’s alpaca models, and generate the data.
\newblock \url{https://github.com/tatsu-lab/stanford_alpaca}.

\bibitem[Thakkar et~al.(2023)Thakkar, Ramani, Cecka, Shivam, Lu, Yan, Kosaian, Hoemmen, Wu, Kerr, Nicely, Merrill, Blasig, Qiao, Majcher, Springer, Hohnerbach, Wang, and Gupta]{Thakkar_CUTLASS_2023}
Vijay Thakkar, Pradeep Ramani, Cris Cecka, Aniket Shivam, Honghao Lu, Ethan Yan, Jack Kosaian, Mark Hoemmen, Haicheng Wu, Andrew Kerr, Matt Nicely, Duane Merrill, Dustyn Blasig, Fengqi Qiao, Piotr Majcher, Paul Springer, Markus Hohnerbach, Jin Wang, and Manish Gupta.
\newblock {CUTLASS}, January 2023.
\newblock URL \url{https://github.com/NVIDIA/cutlass}.

\bibitem[Touvron et~al.(2023)Touvron, Lavril, Izacard, Martinet, Lachaux, Lacroix, Rozière, Goyal, Hambro, Azhar, Rodriguez, Joulin, Grave, and Lample]{touvron2023llamaopenefficientfoundation}
Hugo Touvron, Thibaut Lavril, Gautier Izacard, Xavier Martinet, Marie-Anne Lachaux, Timothée Lacroix, Baptiste Rozière, Naman Goyal, Eric Hambro, Faisal Azhar, Aurelien Rodriguez, Armand Joulin, Edouard Grave, and Guillaume Lample.
\newblock Llama: Open and efficient foundation language models, 2023.
\newblock URL \url{https://arxiv.org/abs/2302.13971}.

\bibitem[Vaswani et~al.(2023)Vaswani, Shazeer, Parmar, Uszkoreit, Jones, Gomez, Kaiser, and Polosukhin]{vaswani2023attentionneedtransformer}
Ashish Vaswani, Noam Shazeer, Niki Parmar, Jakob Uszkoreit, Llion Jones, Aidan~N. Gomez, Lukasz Kaiser, and Illia Polosukhin.
\newblock Attention is all you need, 2023.
\newblock URL \url{https://arxiv.org/abs/1706.03762}.

\bibitem[Wei et~al.(2023)Wei, Wang, Schuurmans, Bosma, Ichter, Xia, Chi, Le, and Zhou]{wei2023chainofthoughtpromptingelicitsreasoning}
Jason Wei, Xuezhi Wang, Dale Schuurmans, Maarten Bosma, Brian Ichter, Fei Xia, Ed~Chi, Quoc Le, and Denny Zhou.
\newblock Chain-of-thought prompting elicits reasoning in large language models, 2023.
\newblock URL \url{https://arxiv.org/abs/2201.11903}.

\bibitem[Xiao et~al.(2024)Xiao, Lin, Seznec, Wu, Demouth, and Han]{xiao2024smoothquantaccurateefficientposttraining}
Guangxuan Xiao, Ji~Lin, Mickael Seznec, Hao Wu, Julien Demouth, and Song Han.
\newblock Smoothquant: Accurate and efficient post-training quantization for large language models, 2024.
\newblock URL \url{https://arxiv.org/abs/2211.10438}.

\bibitem[Xiong et~al.(2024)Xiong, Bian, Li, Li, Du, Wang, Yin, and Helal]{search1_xiong2024searchengineservicesmeet}
Haoyi Xiong, Jiang Bian, Yuchen Li, Xuhong Li, Mengnan Du, Shuaiqiang Wang, Dawei Yin, and Sumi Helal.
\newblock When search engine services meet large language models: Visions and challenges, 2024.
\newblock URL \url{https://arxiv.org/abs/2407.00128}.

\bibitem[Yang et~al.(2024)Yang, Yang, Hui, Zheng, Yu, Zhou, Li, Li, Liu, Huang, Dong, Wei, Lin, Tang, Wang, Yang, Tu, Zhang, Ma, Yang, Xu, Zhou, Bai, He, Lin, Dang, Lu, Chen, Yang, Li, Xue, Ni, Zhang, Wang, Peng, Men, Gao, Lin, Wang, Bai, Tan, Zhu, Li, Liu, Ge, Deng, Zhou, Ren, Zhang, Wei, Ren, Liu, Fan, Yao, Zhang, Wan, Chu, Liu, Cui, Zhang, Guo, and Fan]{yang2024qwen2technicalreport}
An~Yang, Baosong Yang, Binyuan Hui, Bo~Zheng, Bowen Yu, Chang Zhou, Chengpeng Li, Chengyuan Li, Dayiheng Liu, Fei Huang, Guanting Dong, Haoran Wei, Huan Lin, Jialong Tang, Jialin Wang, Jian Yang, Jianhong Tu, Jianwei Zhang, Jianxin Ma, Jianxin Yang, Jin Xu, Jingren Zhou, Jinze Bai, Jinzheng He, Junyang Lin, Kai Dang, Keming Lu, Keqin Chen, Kexin Yang, Mei Li, Mingfeng Xue, Na~Ni, Pei Zhang, Peng Wang, Ru~Peng, Rui Men, Ruize Gao, Runji Lin, Shijie Wang, Shuai Bai, Sinan Tan, Tianhang Zhu, Tianhao Li, Tianyu Liu, Wenbin Ge, Xiaodong Deng, Xiaohuan Zhou, Xingzhang Ren, Xinyu Zhang, Xipin Wei, Xuancheng Ren, Xuejing Liu, Yang Fan, Yang Yao, Yichang Zhang, Yu~Wan, Yunfei Chu, Yuqiong Liu, Zeyu Cui, Zhenru Zhang, Zhifang Guo, and Zhihao Fan.
\newblock Qwen2 technical report, 2024.
\newblock URL \url{https://arxiv.org/abs/2407.10671}.

\bibitem[Yang et~al.(2023)Yang, Chen, Jiang, Cho, Huang, and Lu]{yang2023palrpersonalizationawarellms}
Fan Yang, Zheng Chen, Ziyan Jiang, Eunah Cho, Xiaojiang Huang, and Yanbin Lu.
\newblock Palr: Personalization aware llms for recommendation, 2023.
\newblock URL \url{https://arxiv.org/abs/2305.07622}.

\bibitem[Yu et~al.(2022)Yu, Jeong, Kim, Kim, and Chun]{orca}
Gyeong-In Yu, Joo~Seong Jeong, Geon-Woo Kim, Soojeong Kim, and Byung-Gon Chun.
\newblock Orca: A distributed serving system for {Transformer-Based} generative models.
\newblock In \emph{16th USENIX Symposium on Operating Systems Design and Implementation (OSDI 22)}, pages 521--538, Carlsbad, CA, July 2022. USENIX Association.
\newblock ISBN 978-1-939133-28-1.
\newblock URL \url{https://www.usenix.org/conference/osdi22/presentation/yu}.

\bibitem[Zhang et~al.(2024)Zhang, Tang, Song, Meng, Qian, Shao, Song, Zhu, and Gu]{robot2}
Jiatao Zhang, Lanling Tang, Yufan Song, Qiwei Meng, Haofu Qian, Jun Shao, Wei Song, Shiqiang Zhu, and Jason Gu.
\newblock Fltrnn: Faithful long-horizon task planning for robotics with large language models.
\newblock In \emph{2024 IEEE International Conference on Robotics and Automation (ICRA)}, pages 6680--6686, 2024.
\newblock \doi{10.1109/ICRA57147.2024.10611663}.

\bibitem[Zhao et~al.(2024)Zhao, Israel, den Broeck, and Grover]{zhao2024prepackingsimplemethodfast}
Siyan Zhao, Daniel Israel, Guy~Van den Broeck, and Aditya Grover.
\newblock Prepacking: A simple method for fast prefilling and increased throughput in large language models, 2024.
\newblock URL \url{https://arxiv.org/abs/2404.09529}.

\bibitem[Zheng et~al.(2023)Zheng, Chiang, Sheng, Zhuang, Wu, Zhuang, Lin, Li, Li, Xing, Zhang, Gonzalez, and Stoica]{zheng2023judgingllmasajudgemtbenchchatbot}
Lianmin Zheng, Wei-Lin Chiang, Ying Sheng, Siyuan Zhuang, Zhanghao Wu, Yonghao Zhuang, Zi~Lin, Zhuohan Li, Dacheng Li, Eric~P. Xing, Hao Zhang, Joseph~E. Gonzalez, and Ion Stoica.
\newblock Judging llm-as-a-judge with mt-bench and chatbot arena, 2023.
\newblock URL \url{https://arxiv.org/abs/2306.05685}.

\bibitem[Zheng et~al.(2024)Zheng, Yin, Xie, Sun, Huang, Yu, Cao, Kozyrakis, Stoica, Gonzalez, Barrett, and Sheng]{zheng2024sglangefficientexecutionstructured}
Lianmin Zheng, Liangsheng Yin, Zhiqiang Xie, Chuyue Sun, Jeff Huang, Cody~Hao Yu, Shiyi Cao, Christos Kozyrakis, Ion Stoica, Joseph~E. Gonzalez, Clark Barrett, and Ying Sheng.
\newblock Sglang: Efficient execution of structured language model programs, 2024.
\newblock URL \url{https://arxiv.org/abs/2312.07104}.

\bibitem[Zhong et~al.(2024)Zhong, Liu, Chen, Hu, Zhu, Liu, Jin, and Zhang]{distserve}
Yinmin Zhong, Shengyu Liu, Junda Chen, Jianbo Hu, Yibo Zhu, Xuanzhe Liu, Xin Jin, and Hao Zhang.
\newblock {DistServe}: Disaggregating prefill and decoding for goodput-optimized large language model serving.
\newblock In \emph{18th USENIX Symposium on Operating Systems Design and Implementation (OSDI 24)}, pages 193--210, Santa Clara, CA, July 2024. USENIX Association.
\newblock ISBN 978-1-939133-40-3.
\newblock URL \url{https://www.usenix.org/conference/osdi24/presentation/zhong-yinmin}.

\bibitem[Zhou et~al.(2025)Zhou, Su, Chi, Zhang, Wang, Huang, Sheng, and Wang]{robot1hou2025codeasmonitorconstraintawarevisualprogramming}
Enshen Zhou, Qi~Su, Cheng Chi, Zhizheng Zhang, Zhongyuan Wang, Tiejun Huang, Lu~Sheng, and He~Wang.
\newblock Code-as-monitor: Constraint-aware visual programming for reactive and proactive robotic failure detection, 2025.
\newblock URL \url{https://arxiv.org/abs/2412.04455}.

\end{thebibliography}

\clearpage

% \beginappendix

% \input{sections/appendix}

\end{document}